\documentclass[12pt]{iopart}
\usepackage{graphicx}
\usepackage{color}
\newcommand{\vomega}{\mbox{\boldmath $ \omega $}}
\newcommand{\vOmega}{\mbox{\boldmath $ \Omega $}}
\newcommand{\vmu}{\mbox{\boldmath $ \mu $}}
\begin{document}

\title[Hydrodynamics and magnetohydrodynamics inside a rotating sphere]
  {Hydrodynamic and magnetohydrodynamic computations inside a rotating sphere}
\author{P D Mininni$^1$, D C Montgomery$^2$, and L Turner$^3$}
\address{$^1$ National Center for Atmospheric Research, P.O. Box 3000, 
  Boulder, CO 80307, USA}
\address{$^2$ Dept. of Physics and Astronomy, Dartmouth College, Hanover, 
  NH 03755, USA}
\address{$^3$ Dept. of Astronomy, Cornell University, Ithaca, NY 14853, USA}
\ead{mininni@ucar.edu}

\begin{abstract}
Numerical solutions of the incompressible magnetohydrodynamic (MHD) 
equations are reported for the interior of a rotating, perfectly-conducting, 
rigid spherical shell that is insulator-coated on the inside. A 
previously-reported spectral method is used which relies on a Galerkin 
expansion in Chandrasekhar-Kendall vector eigenfunctions of the curl. The 
new ingredient in this set of computations is the rigid rotation of the 
sphere. After a few purely hydrodynamic examples are sampled (spin down, 
Ekman pumping, inertial waves), attention is focused on selective 
decay and the MHD dynamo problem. In dynamo runs, prescribed 
mechanical forcing excites a persistent velocity field, usually turbulent 
at modest Reynolds numbers, which in turn amplifies a small seed magnetic 
field that is introduced. A wide variety of dynamo activity is observed, 
all at unit magnetic Prandtl number. The code lacks the resolution to 
probe high Reynolds numbers, but nevertheless interesting dynamo regimes 
turn out to be plentiful in those parts of parameter space in which the 
code is accurate. The key control parameters seem to be mechanical and 
magnetic Reynolds numbers, the Rossby and Ekman numbers (which in our 
computations are varied mostly by varying the rate of rotation of the 
sphere) and the amount of mechanical helicity injected. Magnetic energy 
levels and magnetic dipole behavior are exhibited which fluctuate strongly 
on a time scale of a few eddy turnover times. These seem to stabilize as 
the rotation rate is increased until the limit of the code resolution is 
reached.
\end{abstract}
\submitto{\NJP}
\maketitle

\section{Introduction \label{sec:intro}}
In a previous paper, a spectral method for computing incompressible fluid 
and magnetohydrodynamic (MHD) behavior inside a sphere was introduced 
(Ref. \cite{Mininni06}, hereafter referred to as ``MM''). The emphasis in 
MM was on accurate computation of global flow patterns throughout the full 
sphere (including the origin) with special attention to the computation of 
dynamo action, whereby the kinetic energy of a turbulent conducting fluid 
may give rise to macroscopic magnetic fields. The computations were limited 
to moderate Reynolds numbers, with boundary conditions in which the normal 
components of the velocity field, magnetic field, vorticity, and electric 
current density were required to vanish at a rigid, spherical, 
insulator-lined, perfectly-conducting shell, and the three components of 
the velocity and magnetic field were regular at the origin. The feature 
previously lacking that we wish to explore in the present paper is that of 
uniform rotation of the sphere. We postpone to the future the investigation 
of insulating spherical shells which permit the magnetic field to penetrate 
into a non-conducting region outside \cite{Kono02}, though we remark later 
on some considerations relevant to this modification.

In Section \ref{sec:equations}, we formulate the equations to be solved for 
a uniform-density conducting fluid inside a sphere in a familiar set of 
dimensionless (``Alfvenic'') units. We refer to MM for background and 
such of the details as remain unchanged. The main changes reported here 
are: (1) the introduction of a Coriolis term in the equation of motion 
(the centrifugal term may be absorbed in the pressure for incompressible 
flow); and (2) the velocity field ${\bf v}$ (instead of the vorticity 
$\vomega$) and magnetic field ${\bf B}$ are expanded in orthonormal 
Chandrasekhar-Kendall (``C-K'') vector eigenfunctions of the curl 
\cite{Chandrasekhar57,Montgomery78,Turner83,Cantarella00,Mininni06}. The 
physical situation being simulated is again a perfectly conducting, 
mechanically impenetrable sphere coated on the inside with a thin layer 
of insulator, but now viewed from a coordinate system that is regarded as 
rigidly rotating with the bounding spherical shell. Unsurprisingly, the 
dynamical phenomena resulting are markedly different from, and richer 
than, they were in the case without rotation.

Section \ref{sec:hdresults} describes the results of some purely hydrodynamic 
runs (the code may be readily converted into a Navier-Stokes code by simply 
deleting the terms associated with the magnetic field). Included are 
examples \cite{Greenspan,Acheson} of: (i) spin down, or decay of  
relatively rotating kinetic energy due to the action of viscosity; (ii) 
Ekman pumping with flow patterns that result from rotating boundaries; 
(iii) internal waves, three-dimensional relatives of meteorological Rossby 
waves, that depend on the stabilization introduced by rotation for their 
oscillatory features; and (iv) some mechanically-forced runs with a finite 
angle between the symmetry axis of the forcing and the axis of rotation. 
This fourth case results in columnar vortices due to the effect of rotation 
\cite{Acheson,Zhang00} (note convection is absent in our present formulation).

Section \ref{sec:mhdresults} proceeds to a consideration of the MHD case 
with an emphasis on selective decay and the kinds of dynamo behavior 
we have been able to resolve. The spectral method we use involves 
inherently less resolution than some other methods in use, and we have 
been careful to study parameter regimes only where we can resolve the 
relevant length scales. Selective decay is observed to be somewhat 
arrested as the rotation rate is increased. A pleasant surprise has 
been the wide variety of dynamo behavior we have been able to resolve 
without the need to reach parameter regimes regarded as realistic for 
planetary dynamos 
\cite{Glatzmaier95,Glatzmaier96,Zhang00,Christensen01,Roberts01,Kono02}. 
In the summary, Section \ref{sec:future}, we describe briefly some plans 
we have for improving the resolution of the code by some pseudo-spectral 
modifications and some intended future diversification of the boundary 
conditions.

\section{Computational method \label{sec:equations}}
We begin from the magnetohydrodynamic (MHD) equation of motion in a 
rotating coordinate frame \cite{Moffatt,Kono02},
\begin{equation}
\frac{\partial {\bf v}}{\partial t} = {\bf v} \times \vomega + 
    {\bf j} \times {\bf B} - \nabla \left({\mathcal P} + \frac{v^2}{2} 
    \right) - 2 \vOmega \times {\bf v} + \nu \nabla^2 {\bf v} + {\bf f} , 
\label{eq:momentum}
\end{equation}
and the MHD induction equation,
\begin{equation}
\frac{\partial {\bf B}}{\partial t} = \nabla \times \left( {\bf v} 
    \times {\bf B} \right) + \eta \nabla^2 {\bf B} .
\label{eq:induction}
\end{equation}
In the dimensionless Alfvenic units \cite{Mininni06}, ${\bf v}$ is the 
vector velocity field, ${\bf B}$ is the magnetic induction, 
$ \vomega = \nabla \times {\bf v}$ is the vorticity, and 
${\bf j} = \nabla \times {\bf B}$ is the electric current density. 
The generalized pressure is ${\mathcal P}$. The dimensionless viscosity, 
which, in the dimensionless variables, can be interpreted as the reciprocal 
of a mechanical Reynolds number, is $\nu$, and the magnetic diffusivity, 
which can be interpreted as the reciprocal of a magnetic Reynolds number, 
is $\eta$.  The vector field ${\bf f}$ is a solenoidal, externally-applied 
forcing field which is intended to mimic the presence of mechanical sources 
of excitation of ${\bf v}$. Equations (\ref{eq:momentum}) and 
(\ref{eq:induction}) are to be supplemented by the requirements that the 
divergences of both ${\bf v}$ and ${\bf B}$ must vanish everywhere. 
Dropping equation (\ref{eq:induction}) and the terms in equation 
(\ref{eq:momentum}) containing ${\bf j}$ and ${\bf B}$ leaves the forced 
Navier-Stokes equation, and dropping ${\bf f}$ leaves the unforced version 
of Navier-Stokes. $\Omega$ is the (constant) rotation speed of the 
coordinate system, understood to be attached to a rotating spherical 
shell that constitutes the boundary and is both mechanically impenetrable 
and perfectly conducting, with a thin layer of insulator on the inside 
surface.

The non-trivial boundary conditions imposed are that the normal components 
of ${\bf v}$, ${\bf B}$, ${\bf j}$, and $\vomega$ shall all vanish at the 
radius of a unit sphere centered at the origin. The three components of the 
fields ${\bf v}$ and ${\bf B}$ are also required to be regular at the origin. 
The vanishing of the normal components of ${\bf v}$ and $\vomega$ at the 
surface of the unit sphere are implied by, but do not imply, no-slip 
boundary conditions at that radius. Going further with an attempt to 
implement fully a set of no-slip boundary conditions raises unresolved 
paradoxes with respect to the pressure determination which we prefer not 
to confront here (see Refs. \cite{Kress00,Gallavotti,Mininni06} for a 
discussion of these), believing that their seriousness and intractability 
require consideration in the context of simpler situations than the 
present one.

The spectral technique implemented involves expanding ${\bf v}$ and ${\bf B}$ 
in terms of C-K functions (defined below):
\begin{equation}
{\bf v}({\bf r},t) = \sum_{qlm} \xi^v_{qlm}(t) \, {\bf J}_{qlm}({\bf r}) ,
\label{eq:expanv}
\end{equation}
and
\begin{equation}
{\bf B}({\bf r},t) = \sum_{qlm} \xi^B_{qlm}(t) \, {\bf J}_{qlm}({\bf r}) .
\label{eq:expanb}
\end{equation}

The C-K functions 
\cite{Chandrasekhar57,Montgomery78,Turner83,Cantarella00,Mininni06} 
${\bf J}_i$ are defined by
\begin{equation}
{\bf J}_i = \lambda \nabla \times {\bf r} \psi_i + \nabla \times \left( 
    \nabla \times {\bf r} \psi_i \right) ,
\end{equation}
where we work with a set of spherical orthonormal unit vectors 
$(\hat{r},\hat{\theta},\hat{\phi})$ and the scalar function $\psi_i$ is a 
solution of the Helmholtz equation, $(\nabla^2 + \lambda^2) \psi_i = 0$. 
The explicit form of $\psi_i$ is
\begin{equation}
\psi_i (r, \theta, \phi) = C_{ql} \, j_l(|\lambda_{ql}| r) Y_{lm} 
    (\theta,\phi) ,
\label{eq:psi}
\end{equation}
where $j_l(|\lambda_{ql}| r)$ is a spherical Bessel function of the first 
kind which vanishes at $r=1$ and $Y_{lm}(\theta,\phi)$ is a spherical 
harmonic in the polar angle $\theta$ and the azimuthal angle $\phi$. The 
subindex $i$ is a shorthand notation for the three indices $(q,l,m)$, where 
$q$ indexes the successive values of $\lambda$ that make $j_l$ vanish at 
$r=1$ for each value of $l$; $q=1,2,3,\dots$ corresponds to the positive 
values of $\lambda$, and $q=-1,-2,-3,\dots$ indexes the negative values; 
finally $l=1,2,3,\dots$ and $m$ runs in integer steps from $-l$ to $+l$. 
The vectors ${\bf J}_i$ satisfy 
\begin{equation}
\nabla \times {\bf J}_i = \lambda_i {\bf J}_i ,
\end{equation}
and with the proper normalization constants are an orthonormal set that 
has been shown to be complete \cite{Cantarella00}. The integral relation 
expressing the orthogonality of the ${\bf J}_i$ is:
\begin{equation}
\int {\bf J}_{qlm} \cdot {\bf J}_{q',l',m'}^* dV = \delta_{q,q'} 
    \delta_{l,l'} \delta_{m,m'} ,
\end{equation}
where the asterisk denotes complex conjugate, and with the normalization 
constants given by:
\begin{equation}
C_{ql} =  \left|\lambda_{ql} \, j_{l+1}(|\lambda_{ql}|) \right|^{-1} 
    \left[ l(l+1) \right]^{-1/2} .
\label{eq:normalization}
\end{equation}

The scheme for solving equations (\ref{eq:momentum}) and (\ref{eq:induction}) 
is conceptually simple. We substitute the expansions (\ref{eq:expanv}) 
and (\ref{eq:expanb}) into equations (\ref{eq:momentum}) and 
(\ref{eq:induction}), utilize the fact that the ${\bf J}_i$ are 
eigenfunctions of the curl, and then take inner products one at a 
time with the individual ${\bf J}_i$. Their orthogonality enables us 
to pick off expressions  for the time derivatives of the time dependent 
expansion coefficients $\xi_i^v$ and $\xi_i^B$, and equations 
(\ref{eq:momentum}) and (\ref{eq:induction}) are thereby converted 
into a set of ordinary differential equations for the expansion 
coefficients. These appear as 
\begin{equation}
\frac{\partial \xi^v_i}{\partial t} = \sum_{j,k} A^i_{jk} \left( 
    \xi^v_j \xi^v_k - \xi^B_j \xi^B_k \right) + 
    2 \sum_j \vOmega \cdot {\bf O}^i_j - \nu \lambda_i^2 
    \xi^v_i + \xi^f_i ,
\label{eq:CK1}
\end{equation}
and
\begin{equation}
\frac{\partial \xi^B_i}{\partial t} = \sum_{j,k} B^i_{jk} \xi^v_j \xi^B_k
    - \eta \lambda_i^2 \xi^B_i ,
\label{eq:CK2}
\end{equation}
with the coupling coefficients defined as
\begin{equation}
A^i_{jk} = \lambda_k I^i_{jk} , \;\;\;\;\; B^i_{jk} = \lambda_i I^i_{jk} ,
\end{equation}
\begin{equation}
I^i_{jk} = \int {\bf J}_i^* \cdot {\bf J}_j \times {\bf J}_k dV , \;\;\;\;\; 
{\bf O}^i_j = \int {\bf J}_i^* \times {\bf J}_j dV .
\end{equation}

\begin{figure}
\begin{center}\includegraphics[width=15cm]{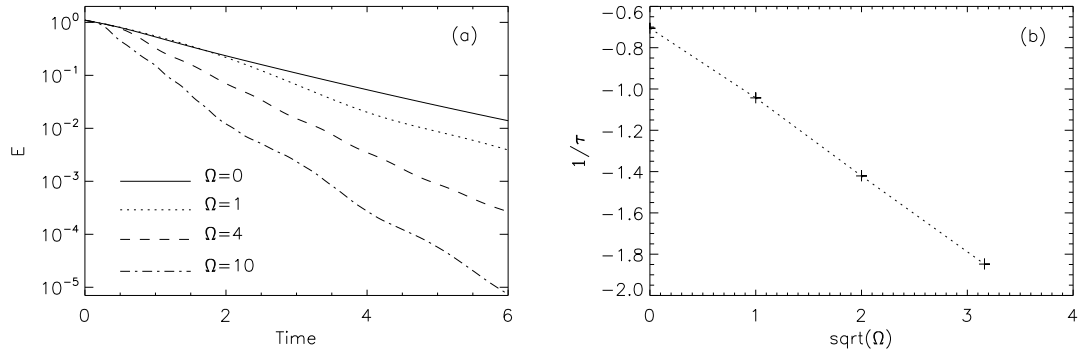}
\end{center}
\caption{(a) Time histories of  the decay of mechanical energy for four 
hydrodynamic runs (H1 to H4) with identical initial velocity fields but 
different rotation rates $\Omega$. (b) Decay rates vs. the square root 
of $\Omega$ for the same four runs.}
\label{fig:spin}
\end{figure}

The infinite set of ordinary differential equations is truncated at some 
level above maximum values of $|q|$ and $l$, in the usual manner of a 
Galerkin approximation \cite{Canuto}. The evaluation of equations 
(\ref{eq:CK1}) and (\ref{eq:CK2}) and the storage of the resulting 
arrays of coupling coefficients in tables, are the most demanding 
numerical tasks of the problem. Once available, they do not have to 
be recomputed, and provide a method for verifying the ideal quadratic 
conservation laws with high accuracy \cite{Mininni06}. Also, since 
the method is purely spectral and fields are only computed in 
real space for visualization purposes, there is no numerical 
singularity in the center of the sphere.

The main drawback of the scheme, as with any wholly spectral one, is that 
the convolution sums in equations (\ref{eq:CK1}) and (\ref{eq:CK2}) grow 
rapidly with increasing maximum values of $|q|$ and $l$, and limit the 
resolution when compared to pseudospectral computations utilizing fast 
transforms (in practice, a resolution of $\max \{|q|\} = \max \{l\} = 9$ 
was used in all the runs). This limits us to modest Reynolds numbers 
(all our computations reported here have limited themselves to resolvable 
Reynolds numbers). Future plans include pseudospectral modifications to 
the evaluation of at least the angular parts of the nonlinear terms in 
equations (\ref{eq:momentum}) and (\ref{eq:induction}), as will be 
mentioned again in the final summary (Section \ref{sec:future}).

\section{Hydrodynamic examples \label{sec:hdresults}}

Some neutral-fluid effects (good introductions to all of which may be 
found in Refs. \cite{Greenspan,Acheson}) are treated first before proceeding 
to MHD. It is worth noticing here that although our boundary conditions 
are implied by, but do not imply no-slip velocities, several qualitative 
and some quantitative agreements are observed with previous experiments 
and theory. First we study simple problems in which initially relatively 
rotating fluids adjust themselves to rigid rotation with the spherical 
shell. We study these decays as functions of the rotation rate $\Omega$. 
The prediction \cite{Greenspan} is that the decay of the non-rigid body 
components should be exponential, with a decay rate that varies as 
$\Omega^{1/2}$. Figure \ref{fig:spin}(a) shows the time histories of the 
decay of mechanical energy for four runs with identical initial velocity 
fields limited to a few random low mode numbers (large spatial scales). 
The runs are delineated as runs H1, H2, H3, and H4. 
Specifically, we have as initial conditions a random superposition of 
modes with $|q|=1,2$, $l=1,2$, and all allowed values of $m$, with 
viscosity $\nu = 0.01$ and values of $\Omega$ of $0$, $1$, $4$, and 
$10$ respectively. Each curve is approximately exponential, and when they 
are fitted with exponentials, the decay rates plotted vs. the square root of 
$\Omega$ appear as in Figure \ref{fig:spin}(b), and are adjudged to be in 
satisfactory agreement with theory \cite{Greenspan}. The nonlinearities 
excite smaller spatial scales, and the decay process is progressively 
enhanced by increasing the rotation rate.

\begin{figure}
\begin{center}\includegraphics[width=16cm]{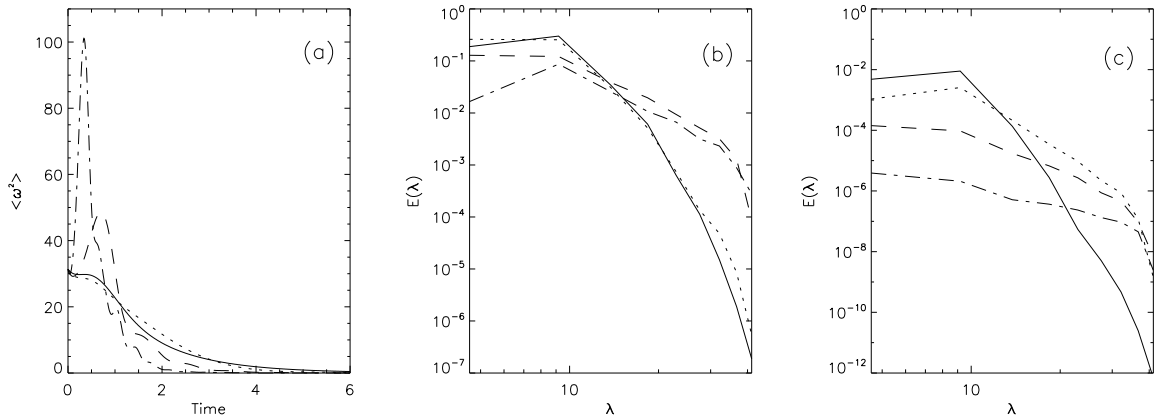}
\end{center}
\caption{(a) Enstrophy for runs H1 to H4 as a function of time (same 
labels as in Figure \ref{fig:spin}). (b) Energy spectra at $t=1$ for 
the same runs, and (c) energy spectra at $t=6$. In all cases, solid 
lines are for $\Omega = 0$, dotted lines for $\Omega =1$, dashed lines 
for $\Omega =4$, and dash-dotted lines for $\Omega =10$.}
\label{fig:ekman}
\end{figure}

A second feature of rotating spherical flows observed in our code is the 
development of Ekman-like layers and the action of Ekman pumping 
\cite{Greenspan,Acheson} (see also \cite{Dormy98} for a detailed study 
in rotating spherical shells). The flow patterns are characterized by the 
development of interior vortical flows with some symmetries, and thin 
layers that separate the large vortices and also lie along the wall 
boundary layers. These have a characteristic thickness of the order of 
$\delta \sim E_K^{1/2} R$, where $R=1$ is the radius of the sphere, and 
the Ekman number is $E_K = \nu \Omega^{-1} L^{-2}$, with $L$ a 
characteristic scale of the flow. The ability of the code to compute 
these layers is limited by its resolution.  Realistic values of the 
Ekman number are, for planetary core regimes 
\cite{Moffatt,Zhang00,Kono02}, beyond our range. In all the runs 
presented here we will limit ourselves to cases where $\delta$ can 
be properly resolved with the number of modes used in the simulations. The 
presence of the Ekman layers is concomitant with the development of smaller 
spatial scales and hence of more rapid dissipation. Figure 
\ref{fig:ekman} illustrates this fact. Figure \ref{fig:ekman}(a) is the 
integrated squared vorticity, or enstrophy, for the four runs whose decay 
has just been seen to be exponential. The largest rotation rate corresponds 
to the curve with the highest early peak in enstrophy spectrum, the second 
highest with the second largest, and so on. Figure \ref{fig:ekman}(b) 
shows the energy spectra at $t=1$ for the four runs and Figure 
\ref{fig:ekman}(c) shows the same energy spectra at $t=6$. In every 
case, the flatter spectra, and hence the shorter wavelength dominances, 
correspond to the higher values of $\Omega$. This is the result of 
the formation of a thinner boundary layer as $\Omega$ is increased.

\begin{figure}
\begin{center}\includegraphics[width=15cm]{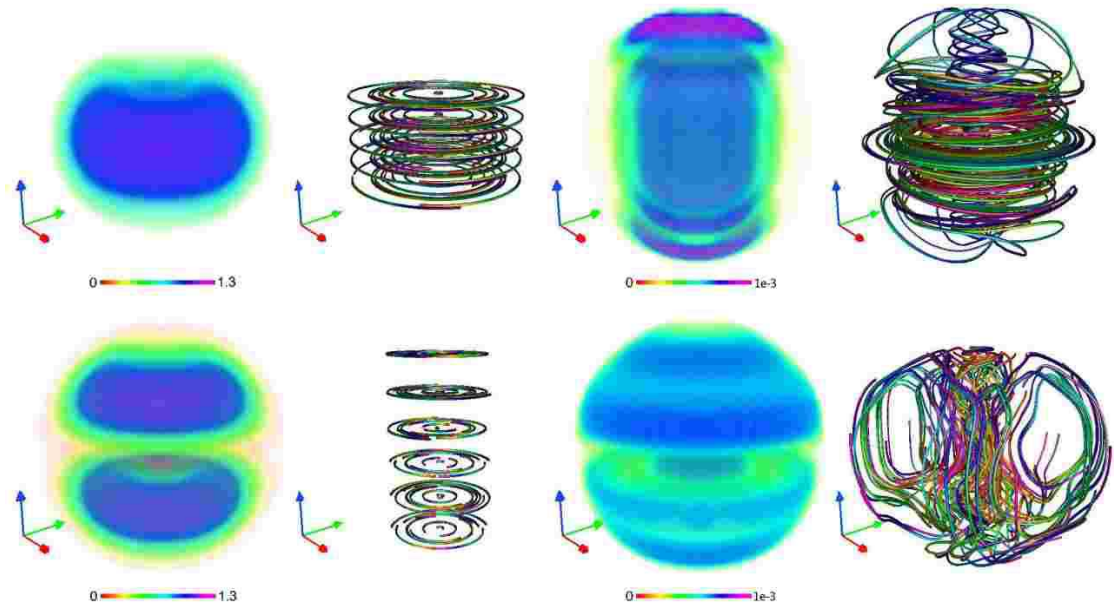}
\end{center}
\caption{Above: Mechanical energy density and velocity field lines in run 
E1, at $t=0$ (left) and at $t=6$ (right). Below: idem for run E2. For 
convenience, energies and field lines are always shown in pairs, with 
energy densities on the left and field lines on the right. The field lines 
change color according to the distance integrated from the initial point, 
from red to yellow, blue, and magenta. The red, green, and blue arrows 
indicate respectively the $x$, $y$, and $z$ axis. $\vOmega$ is in the 
$z$ direction. In both cases, the energy density is symmetric with 
respect to the equator, while the flow itself is antisymmetric in run E1 
(above).}
\label{fig:3Dekman}
\end{figure}

\begin{figure}
\begin{center}\includegraphics[width=13cm]{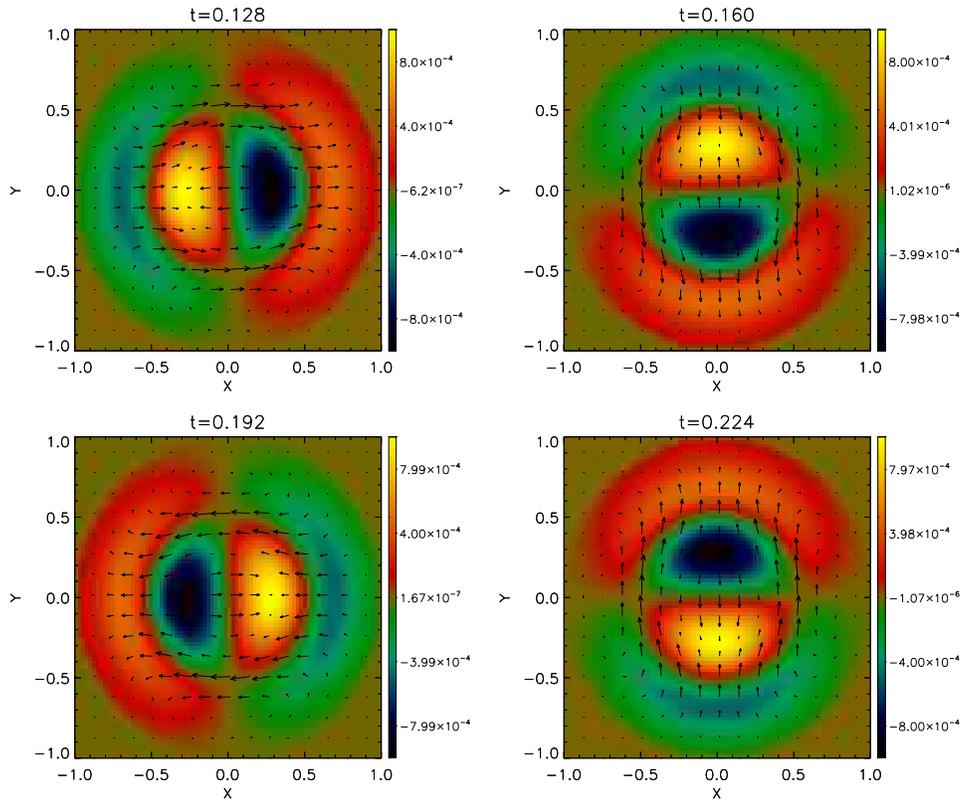}
\end{center}
\caption{Equatorial cross section of the velocity field for an inertial 
wave in the rotating sphere. The reference frame is fixed to the sphere. 
$\vOmega$ is in the $z$ direction. The axial velocity is indicated by 
the colors, while the radial and azimuthal velocities are indicated by 
the arrows. For this mode, the frequency is $\varpi \approx 49$.}
\label{fig:wave}
\end{figure}

The range of Ekman layer behavior we have been able to observe is very wide. 
We show in Figure \ref{fig:3Dekman} the results of two simulations (labeled 
E1 and E2) with initially random axisymmetric ($m=0$) velocity fields which 
are purely azimuthal. For both runs, the initial velocity is proportional to 
the difference between ${\bf J}_{q,l,0}$ and ${\bf J}_{-q,l,0}$, which has 
only an azimuthal component. For run E1, $q=1$ and $l=1$, and for run E2, 
$q=1$ and $l=2$. Both runs have $\nu=0.01$ and $\Omega=10$. The time evolution 
of the runs is similar to the evolution displayed in Figures \ref{fig:spin} 
and \ref{fig:ekman}. However, the axisymmetric initial conditions in runs 
E1 and E2 make visualization of flow patters easier. Figure \ref{fig:3Dekman} 
shows the initial and late-time flow patterns for these runs, using the 
VAPOR graphics package \cite{vapor} that will be repeatedly used throughout 
this paper for graphical demonstrations. The rotation generates poloidal 
components of the velocity field fast, and at late times different patterns 
are observed depending on the initial value of $l$. In run E1, at late 
times the flow displays a poloidal circulation on top of the initial 
toroidal field: the flow is directed towards the center of the sphere 
along the axis of rotation, and a return flow is observed in both 
hemispheres close to the wall. In other words, the flow can be 
described as the superposition of a toroidal differential rotation and a 
poloidal meridional circulation. This circulation is radially outward in 
the meridional plane, directed towards the poles close to the wall, and 
redirected toward the equatorial plane again as the flow gets close to 
the poles. Both hemispheres show the same pattern. In run E2 the pattern 
is more complex, and vertical velocities are observed in the vicinity of 
the axis of rotation, while high and at intermediate latitudes a poloidal 
circulation is generated.

\begin{figure}
\begin{center}\includegraphics[width=14cm]{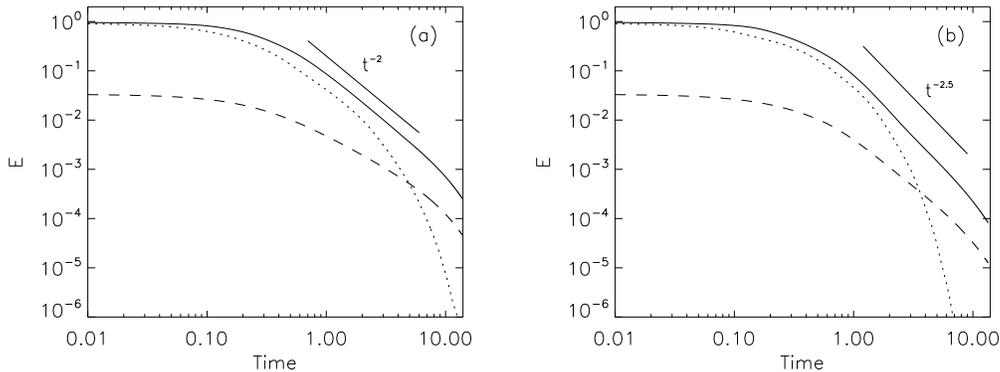}
\end{center}
\caption{(a) Magnetic energy $E_M$ (solid line), kinetic energy $E_V$ 
(dotted line), and magnetic helicity $H_M$ (dashed line) in the selective 
decay run S1. (b) Same quantities for run S2. The two power laws are 
indicated in the figures only as a reference.}
\label{fig:selective1}
\end{figure}

As a third hydrodynamic test of the code we demonstrate inertial wave 
motion in the small-amplitude limit. Equations (\ref{eq:CK1}) and 
(\ref{eq:CK2}) can be linearized in powers of a small departure from a 
uniform rotation velocity; then solutions can be sought which vary with 
time as $e^{i \varpi t}$. The resulting linear homogeneous algebraic 
system can be solved in a Galerkin approximation by expanding the velocity 
and vorticity in the C-K functions. An anti-Hermitian matrix results whose 
eigenfunctions can be found numerically and whose corresponding eigenvalues 
$i \varpi$ may be computed numerically in the process.  Then any one of 
the oscillatory modes can be loaded numerically into the ideal version 
of the code and run with the overall amplitudes chosen to be very small. 
The time evolution is accurately predicted by the computation of the 
single modes, which are standing waves. Figure \ref{fig:wave} shows four 
equatorial cross-sections of the sphere at different times with the axial 
velocity indicated by the color codes, and the radial and azimuthal 
velocities indicated by arrows ($\Omega = 10$ in this run). The times 
are chosen to be one quarter-period apart. The oscillation frequency 
$\varpi$ for the particular mode shown as obtained from the eigenvalue 
problem is in good agreement with the results obtained from the fully 
non-linear code ($\varpi \approx 49$).

In forced hydrodynamic simulations in the presence of strong rotation, 
we also observe the development of columnar structures in the flow, aligned 
with the axis of rotation. The discussion of these simulations will be 
left for the next section, where the connection between these columns and 
dynamo action will be considered.

\section{MHD and the dynamo \label{sec:mhdresults}}

\subsection{Selective decay in the sphere}

\begin{figure}
\begin{center}\includegraphics[width=8cm]{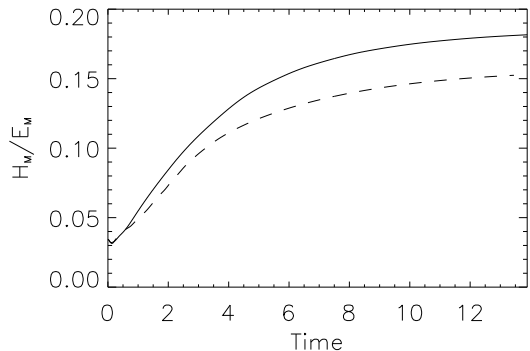}
\end{center}
\caption{Relative helicity $H_M/E_M$ in selective decay runs S1 (solid) 
and S2 (dashed).}
\label{fig:selective2}
\end{figure}

Before passing to a discussion of the mechanically driven spherical dynamo, 
we present first the results of two tests of three-dimensional MHD 
``selective decay'' as affected by the presence of rotation. Selective 
decay is a familiar turbulent decay process, usually incompressible, 
long studied in periodic geometry \cite{Matthaeus80,Ting86,Kinney95}, 
wherein one ideal invariant is cascaded to short wavelengths and 
dissipated while another remains locked into long wavelengths and is 
approximately conserved. The phenomenon, closely connected with inverse 
cascade processes for driven systems, leads toward a state in which 
the ratio of the two ideal invariants involved is minimized and which 
therefore is accessible to variational methods. In 3D MHD, a quantity 
that may be preferentially dissipated is the total energy $E = E_V+E_M$ 
(kinetic plus magnetic) while magnetic helicity $H_M$ may be 
approximately conserved. Under other circumstances, energy may be 
dissipated while cross helicity $K$ is approximately conserved, 
leading to the phenomenon of ``dynamic alignment,'' 
\cite{Grappin83,Pouquet86,Ghosh88} in which the velocity field and 
magnetic fields are highly correlated. The definitions of $H_M$ and 
$K$ are
\begin{equation}
H_M = \frac{1}{2} \int {\bf A} \cdot {\bf B} dV,
\end{equation}
\begin{equation}
K = \frac{1}{2} \int {\bf u} \cdot {\bf B} dV,
\end{equation}
where ${\bf A}$ is the vector potential whose curl is ${\bf B}$, and 
the integrals run over the entire volume of the fluid. What we are 
interested in demonstrating here is the effect that rotation has on 
the development of the selective decay of total energy relative to 
magnetic helicity inside a sphere. It will be useful for these and 
other purposes, to have a definition of the magnetic dipole moment:
\begin{equation}
\vmu = \frac{1}{2} \int {\bf r} \times {\bf j} \, dV \, .
\end{equation}
which is seen to be readily expressible in terms of the expansion 
coefficients for ${\bf B}$ \cite{Mininni06}.

\begin{figure}
\begin{center}\includegraphics[width=7.2cm]{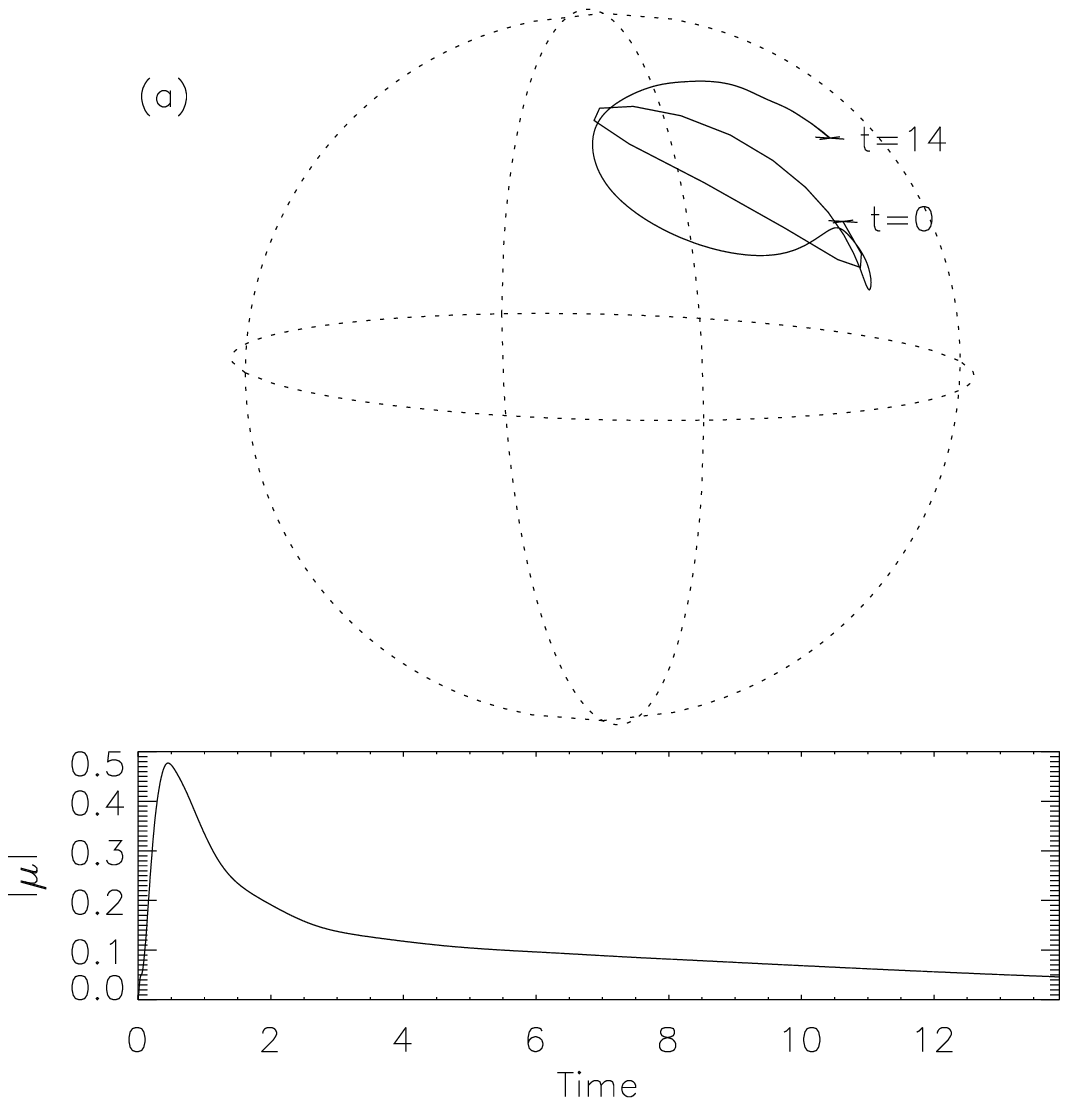}
              \includegraphics[width=7.2cm]{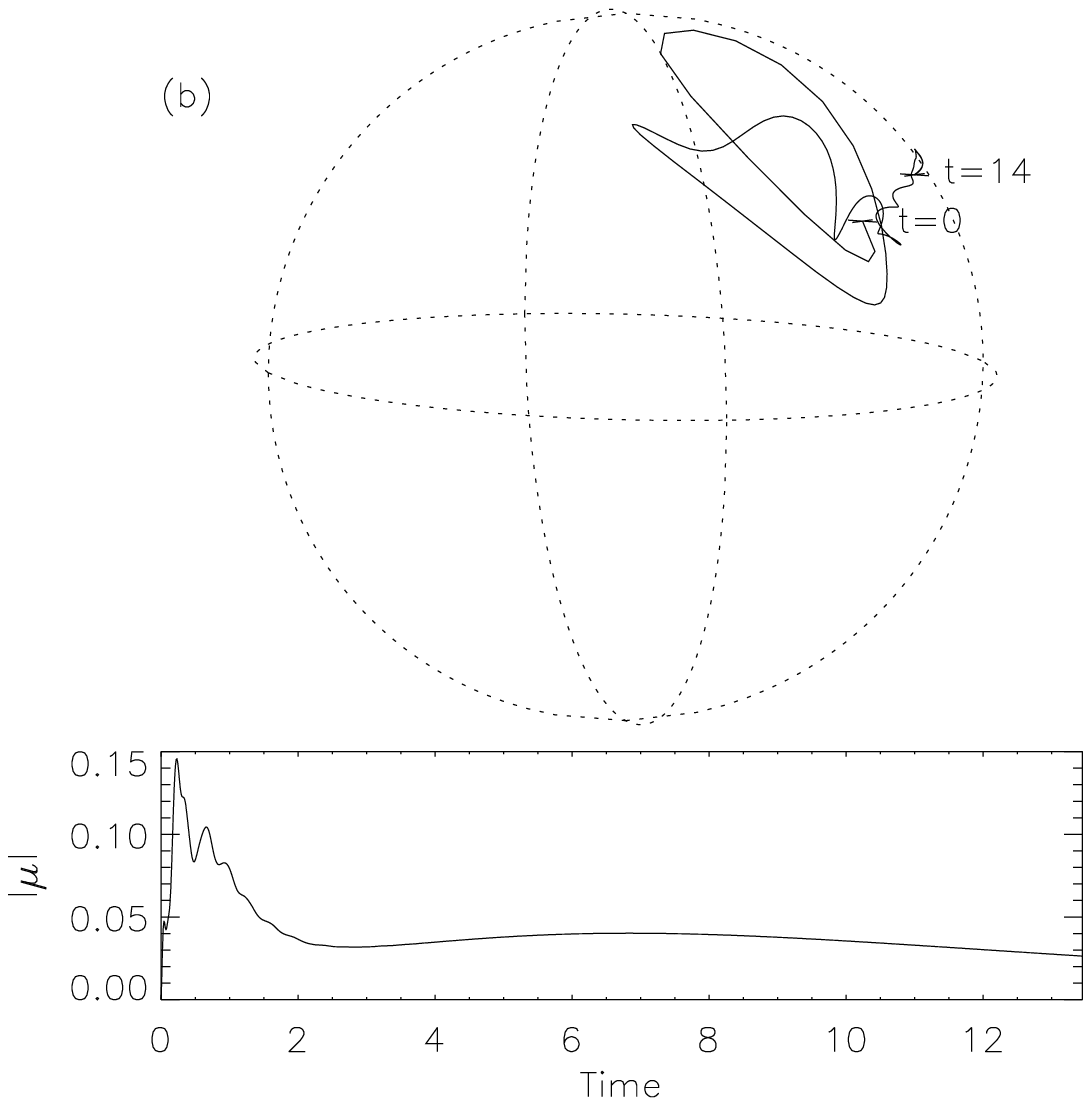}
\end{center}
\caption{(a) Trace of the dipole moment on the surface of the unit sphere 
(above) and amplitude of the dipole moment as a function of time (below) 
for the selective decay run S1. (b) Same quantities for run S2.}
\label{fig:selec_dip}
\end{figure}

The initially excited modes for the two runs we will present 
(``S1'' and ``S2'') are those for $q=\pm 3$, $l=3$, and all possible 
values of $m$. The initial values chosen for the expansion coefficients 
are:
\begin{eqnarray}
& \xi^v_{\pm 3,3,0} = -u_0 , \;\; \xi^v_{\pm 3,3,0<m\le 3} = u_0 (1+i) , \\
& \xi^B_{3,3,0} = \frac{10}{6} \xi^B_{-3,3,0} = b_0 , \;\; 
  \xi^B_{3,3,0<m\le 3} = \frac{10}{6} \xi^B_{-3,3,0<m\le 3} = b_0 (1-i) , 
\end{eqnarray}
with $u_0$ and $b_0$ chosen so that at $t=0$, the magnetic and kinetic 
energies are $E_M=E_V \approx 0.5$, $K=0$, and $H_M \approx 0.034$. Some 
helicity cancellation occurs because of the two signs of $\lambda$ (or $q$). 
As a comparison, note that for the $q=3$, $l=3$ mode alone, $H_M/E_M$ 
is no more than about $0.072$ (this is the maximum value of $|H_M/E_M|$ 
if only modes with $|q|=3$, $l=3$, and one sign of $\lambda$ are excited).
In both runs, the magnetic diffusivity and kinematic viscosity are 
$\nu=\eta=0.006$; the Reynolds numbers are $R_e \approx R_m \approx 170$, 
based on the radius of the sphere. The two runs differ by the values of 
$\Omega$ chosen, which are $2$ and $12$, respectively. These mean that 
the Rossby and Ekman numbers of the two runs are, respectively, 
$R_O=U(\Omega R)^{-1}=0.5$, $E_K=0.003$ for S1 and $R_O=0.083$, 
$E_K=0.0005$ for S2. The decay of magnetic energy, kinetic energy, 
and magnetic helicity for Runs S1 and S2 are shown in Figure 
\ref{fig:selective1}. The behavior in Run S1 is not significantly different 
from the non-rotating case \cite{Mininni06}. Note that in both runs, the 
kinetic energy at late times is negligible, and that magnetic and kinetic 
energies decay faster than the magnetic helicity. However, the decay of 
all these quantities in S2 seems to be faster than in S1.

The relative helicity, $H_M/E_M$, is shown for Runs S1 and S2 in Figure 
\ref{fig:selective2}. It will be seen that the increased rate of rotation 
in S2 has somewhat arrested the selective decay, for reasons not totally 
understood. It may be that the rotation has resulted in sufficient 
two-dimensionalization of the flow \cite{Greenspan,Acheson,Zhang00} 
that the inherently three dimensional nature of the selective decay 
has been compromised. But from Figure \ref{fig:selective2} it can also 
be seen that the ratio of $H_M$ to $E_M$ has approached reasonably 
closely to its maximal value of 
$\min^{-1}\{|\lambda_{ql}|\} \approx 0.22$ (the maximum value of 
$|H_M/E_M|$ when only modes with $|q|=1$, $l=1$, and one sign of 
$\lambda$ are excited). The maximal value would indicate a total 
disappearance of the non-rotational kinetic energy (i.e., a rigid 
rotation), and all the magnetic energy in the largest-scale modes 
(smallest $|\lambda|$) allowed by the boundary conditions: a 
magnetized, ``frozen'' condition.

\begin{figure}
\begin{center}\includegraphics[width=15cm]{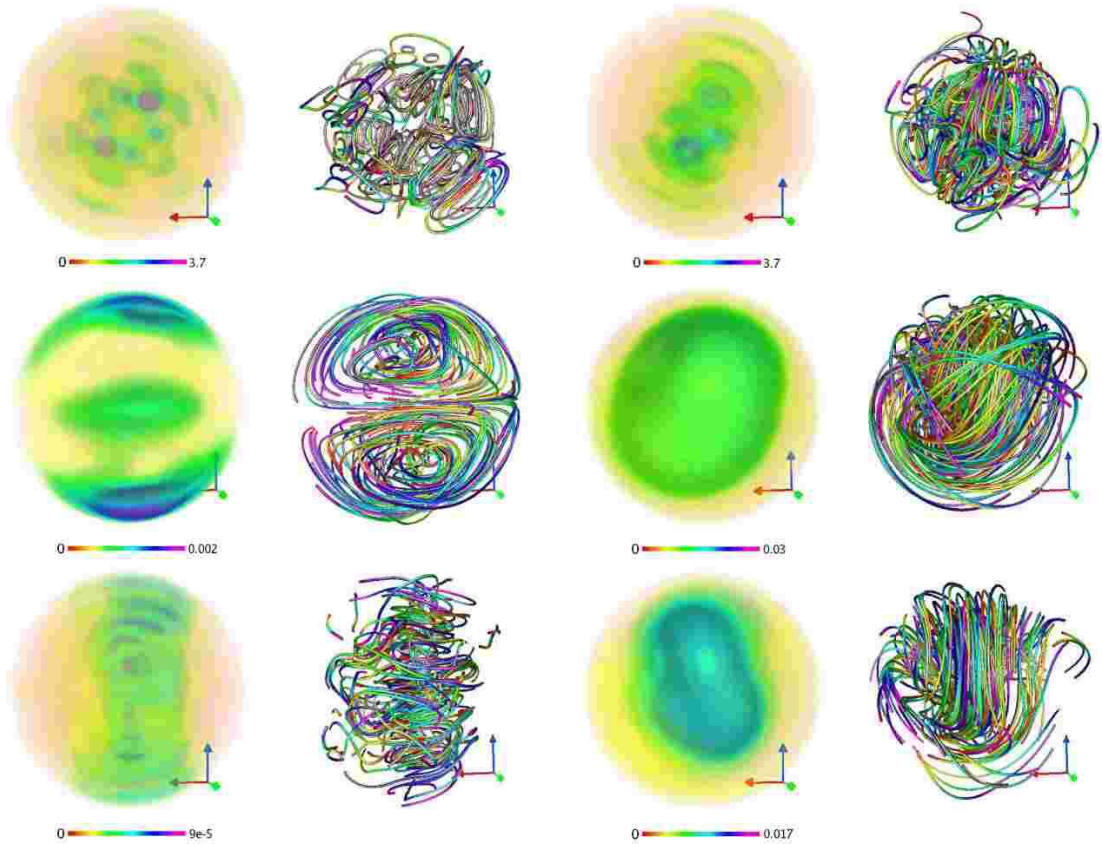}
\end{center}
\caption{Above: kinetic energy density and velocity field lines (left) 
and magnetic energy density and magnetic field lines (right) in the initial 
conditions of selective decay runs S1 and S2. Middle: same fields at $t=14$ 
in run S1. Below: same fields at $t=14$ in run S2. Colors and labels are 
as in Figure \ref{fig:3Dekman}.}
\label{fig:3Dselective}
\end{figure}

The behavior of the dipole moment for the two runs is shown in Figure 
\ref{fig:selec_dip}. The solid lines are the traces of the projected 
direction of the dipole moment on the surface of the sphere as functions 
of the time, and the orientation is such that the axis of rotation points 
upward. In the lower parts of these two figures, the magnitude of the dipole 
moment is plotted vs. time. In both cases, it will be seen that the dipole's 
orientation initially wanders erratically near its initial position, and 
finally ends at a ``mid-latitude'' direction not far from where it began. 
This was something of a surprise to us, since we had expected it to line 
up with the axis of rotation or at least close to it. Figure 
\ref{fig:3Dselective} shows VAPOR plots with the energy densities, velocity 
and magnetic field line structure for the initial conditions in Runs S1 and 
S2, as well as the late stages of both runs ($t=14$) when the selective 
decay process has saturated and all the nonlinear terms are small, preventing 
any further evolution of the system except for dissipation. For strong 
rotation (run S2), the velocity field  is quasi-two dimensional and 
develops column-like structures at late times, while the magnetic field 
is highly anisotropic (although the dipole moment is not aligned with the 
axis of rotation). Note magnetic field lines in this case are aligned 
with the $z$ axis (the axis of rotation), and velocity field lines are 
mostly toroidal. Actually, the ratio $|v_\phi / v_z|$ at $t=14$ averaged 
over the whole volume for this run at $t=14$ is $\approx 13$.

\begin{table}
\caption{Dynamo runs: $q$ and $l$ give the scales where mechanical energy 
is injected by the forcing, $\Omega$ is the rotation rate, and $\nu$ and 
$\eta$ are respectively the kinematic viscosity and magnetic diffusivity. 
``Helical'' indicates whether the forcing injects mechanical helicity, and 
``Axisym.'' indicates whether the forcing is axisymmetric. $\alpha$ is the 
angle between $\vOmega$ and the $z$-axis (the axis of symmetry in 
axisymmetric forcings). Finally, $R_O$, $E_K$, and $R_e$ are respectively 
the Rossby, Ekman, and Reynolds numbers, all based on the radius of the 
sphere. A resolution of $\max \{|q|\} = \max \{l\} = 9$ was used in all 
the runs.}
\label{table:dynamo}
\begin{indented}
\item[]\begin{tabular}{@{}lllllllllll}
\br
Runs & $q$ & $l$ & $\Omega$ & $\nu=\eta$         & Helical & Axisym.
     & $\alpha$  & $R_O$    & $E_K$              & $R_e$ \\
\mr
D1   &  1  &  1  &    2     & $2\times 10^{-3}$  & No      & Yes 
     & $30^\circ$& $0.5$    & $1\times 10^{-3}$  & $500$ \\
D2   &  2  &  2  &    2     & $2\times 10^{-3}$  & No      & Yes  
     & $30^\circ$& $0.5$    & $1\times 10^{-3}$  & $500$ \\
D3   &  3  &  3  &    2     & $2\times 10^{-3}$  & No      & Yes  
     & $30^\circ$& $0.5$    & $1\times 10^{-3}$  & $500$ \\
D4   &  3  &  3  &    4     & $2\times 10^{-3}$  & No      & Yes  
     & $30^\circ$& $0.25$   & $5\times 10^{-4}$  & $500$ \\
D5   &  3  &  3  &    8     & $2\times 10^{-3}$  & No      & Yes  
     & $30^\circ$& $0.125$  & $2.5\times 10^{-4}$& $500$ \\
D6   &  3  &  3  &    8     & $4\times 10^{-3}$  & No      & Yes  
     & $30^\circ$& $0.125$  & $5\times 10^{-4}$  & $250$ \\
D7   &  3  &  3  &    16    & $4\times 10^{-3}$  & No      & Yes  
     & $30^\circ$& $0.0625$ & $2.5\times 10^{-4}$& $250$ \\
D8   &  3  &  3  &    8     & $2\times 10^{-3}$  & No      & Yes  
     & $30^\circ$& $0.125$  & $2.5\times 10^{-4}$& $500$ \\
D9   &  3  &  3  &    16    & $4\times 10^{-3}$  & No      & Yes  
     & $20^\circ$& $0.125$  & $5\times 10^{-4}$  & $250$ \\
D10  &  3  &  3  &    16    & $4\times 10^{-3}$  & No      & No  
     & $0^\circ$ & $0.125$  & $5\times 10^{-4}$  & $250$ \\
D11  &  3  &  3  &    16    & $4\times 10^{-3}$  & Yes     & No 
     & $0^\circ$ & $0.125$  & $5\times 10^{-4}$  & $250$ \\
D12  &  3  &  3  &    16    & $4\times 10^{-3}$  & Yes     & No 
     & $90^\circ$& $0.125$  & $5\times 10^{-4}$  & $250$ \\
\br
\end{tabular}
\end{indented}
\end{table}

\subsection{Dynamos}
We turn now to the case of the forced spherical dynamo computations, in 
which specified non-zero forcing functions ${\bf f}$ are added to the right 
hand side of equation (\ref{eq:momentum}) or (\ref{eq:CK1}) to provide 
a persistently-active, non-decaying velocity field. After a purely 
hydrodynamic run to reach a statistically steady state, very small 
magnetic fields are introduced to see if the velocity fields will cause 
them to amplify, and attention focuses on questions like the orientation 
of the resulting magnetic dipole moment relative to the axis of rotation 
and the dipole moment's magnitude. We are also interested in the kinetic 
and magnetic energy spectra that result, and how the eponymous 
dimensionless numbers of the rotating fluid (Reynolds, magnetic Reynolds, 
Rossby, Ekman) influence the magnetic quantities. These appear to be 
essential control parameters of the problem. The range of possibilities 
is clearly very wide, and we have not begun to explore the entire space 
of possible parameters. Rather, we content ourselves with showing samples 
of different behavior that have emerged for different combinations that 
lead to regimes which the code will resolve satisfactorily. 
Our efforts should not be compared with explorations of the space 
of parameters in realistic geodynamo simulations (see e.g. 
\cite{Christensen06}) but rather as an extension of dynamo simulations 
of incompressible MHD flows (often done using periodic boundary conditions 
\cite{Meneguzzi81,Brandenburg01,Ponty05,Mininni05b}) to include the effect 
of boundaries and rotation.

Several of the forcing functions used in the runs we will display are 
axisymmetric, but their axes of symmetry are not aligned with the axis of 
rotation. The resulting overall asymmetry quickly excites all the available 
retained modes, to some degree. The general form of the axisymmetric forcing 
function used is
\begin{equation}
{\bf f} = \sum_{ql} \xi^f_{q,l,0} \left( A {\bf J}_{q,l,0} - 
    B {\bf J}_{-q,l,0} \right) .
\label{eq:force1}
\end{equation}
For any value of $A$ and $B$, this superposition of C-K functions gives an 
axisymmetric forcing ${\bf f}$. For $A=B=1$, there will be no net helicity 
involved in the forcing (the curl of ${\bf f}$ is perpendicular to 
${\bf f}$), and the only non-vanishing component of velocity that 
is forced is the $\phi$ component (the azimuthal component with respect 
to the axis of symmetry of the forcing, the $z$ axis). This forcing can 
be considered as a simple differential rotation, where the number of nodes 
in $v_\phi (r,\theta)$ is controlled by the values of $q$ and $l$. 
The axis of rotation is typically oriented at some specified angle $\alpha$ 
(often $30^\circ$) to the forcing function's axis of symmetry (the polar 
axis, in spherical coordinates). Thus the rotational motion and the forcing 
have no shared symmetry, and the resulting mechanical motion is totally 
asymmetrical.

\begin{figure}
\begin{center}\includegraphics[width=15cm]{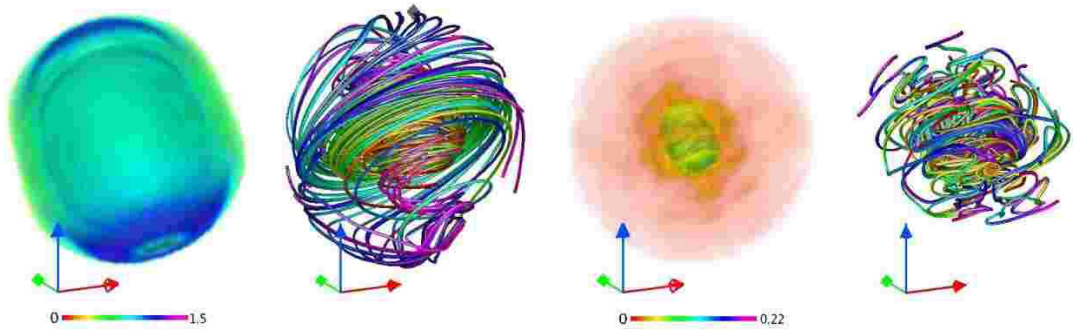}
\end{center}
\caption{Left: kinetic energy density and velocity field lines at late 
times in run D1, when the dynamo has saturated. Right: magnetic energy 
density and magnetic field lines at the same time. Colors and labels are 
as in Figure \ref{fig:3Dekman}.}
\label{fig:D13D}
\end{figure}

\begin{figure}
\begin{center}\includegraphics[width=7.2cm]{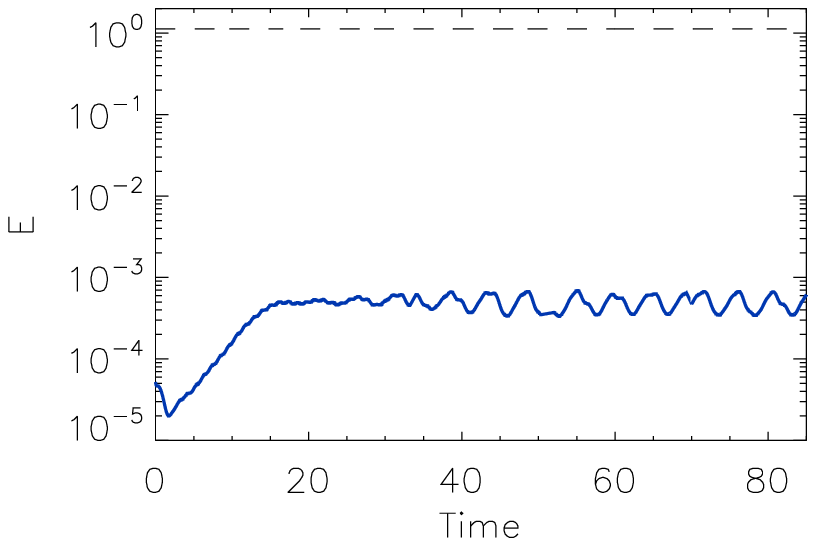}
              \includegraphics[width=7.2cm]{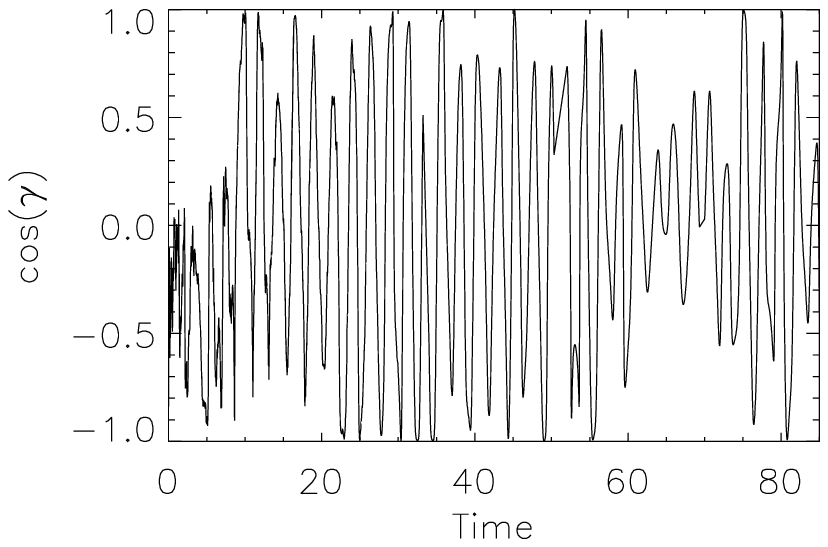}
\end{center}
\caption{Left: kinetic (dashed black line) and magnetic (solid blue line) 
energy as a function of time in run D1. Right: cosine of the angle between 
$\vOmega$ and $\vmu$ in the same run.}
\label{fig:D1evolution}
\end{figure}

Non-axisymmetric forcing functions are obtained by superposing C-K modes 
with $m \neq 0$, i.e.
\begin{equation}
{\bf f} = \sum_{qlm} \xi^f_{qlm} {\bf J}_{qlm} ,
\label{eq:force2}
\end{equation}
and when $\xi^f_{qlm} = \xi^f_{-q,l,m}$ the forcing is non-helical. In 
any other case the curl of ${\bf f}$ has a projection into ${\bf f}$, 
and the forcing injects mechanical helicity into the flow.

In an effort to systematize the runs we have done and the reasons we have 
done them, we have assembled the important parameters for the dynamo runs 
(labeled D1 through D12) in Table \ref{table:dynamo}. Listed in Table 
\ref{table:dynamo} are: the $q$ and $l$ values where the forcing was 
concentrated (determining its characteristic length scale); the rigid 
rotation rate; the kinematic viscosity (reciprocal Reynolds number, if 
the kinetic energy is close to unity) and magnetic diffusivity (this 
study is restricted to the magnetic Prandtl number $P_M = \nu/\eta = 1$ 
case); an indication of whether the forcing was axisymmetric or not; the 
angle between the axis of rotation and the axis of symmetry of the forcing, 
when the forcing function has an internal symmetry; the Rossby and Ekman 
numbers of the flow into which the seed magnetic field is introduced; and 
whether or not the forcing injected net mechanical helicity.

It is perhaps worthwhile to say a word about the motivation for the 
progression of runs shown in Table \ref{table:dynamo}. The first remark is 
that it seems to be relatively easy to excite a dynamo and generate a dipole 
moment, but relatively difficult to generate one that behaves according to 
our predispositions and hopes: a dipole moment with some alignment with the 
axis of rotation, and that reverses periodically or randomly with long times 
between reversals (compared with the turbulent turnover time). We have found 
wild oscillations in both magnitude and direction that seem to decrease with 
decreasing Rossby and Ekman numbers. Since the Reynolds numbers are limited 
by the resolution, the principal means of decreasing both the Rossby and 
Ekman numbers is by increasing the rotation rate $\Omega$. This, however, 
eventually decreases the thickness of the boundary layer below the resolution 
of the code, and beyond that point, the accuracy of the computations becomes 
suspect. The lowest Rossby and Ekman numbers that appear in the entries of 
Table \ref{table:dynamo} represent those below which the resolution 
limitations are encountered. It will be seen below that as we progress 
toward them, the dipolar behavior looks gradually more like what we might 
expect, the dipole moment gets stronger and more aligned with the axis of 
rotation, and the time between reversals gets larger.

We begin by showing some results for a weak dynamo situation, Run D1, 
where the magnetic energy $E_M$ remains always much smaller than the kinetic 
energy $E_V$. In the forcing function (\ref{eq:force1}), the driven modes 
have $q=l=1$, $A=B=1$, $\nu=\eta=0.002$, $\alpha=30^\circ$, and $\Omega=2$. 
The amplitude of the forced modes is $|\xi^f_{q,l,0}|= 0.4$. The two 
Reynolds numbers, $R_e$ and $R_m$, based on the measured r.m.s. velocity 
before the magnetic seed is introduced and the radius of the sphere, are 
both about 500. The Rossby number is $R_O=0.5$, and the Ekman number also 
based on the radius of the sphere is $E_K=1\times 10^{-3}$. Figure 
\ref{fig:D13D} shows the streamlines of the flow, magnetic field lines, and 
the energy densities at late times, once the dipole is established 
($t\approx 80$). In the steady state, the magnetic dipole moment $|\vmu|$ 
is of the order of $0.001$ and the ratio of magnetic to kinetic energies 
is $E_M/E_V \approx 0.0005$.  The magnetic energy rises to a characteristic 
value and oscillates somewhat irregularly as shown in Figure 
\ref{fig:D1evolution}, while the cosine of $\gamma$ (the angle between 
the axis of rotation and the orientation of $\vmu$) oscillates with 
roughly the same periodicity (see also Figure \ref{fig:D1evolution}) 
with almost a $180^\circ$ variation in some cases, and with an average 
orientation almost perpendicular to $\vOmega$. In this weak case D1, 
the hydrodynamic flow remains laminar, stable, and almost time independent.

\begin{figure}
\begin{center}\includegraphics[width=14cm]{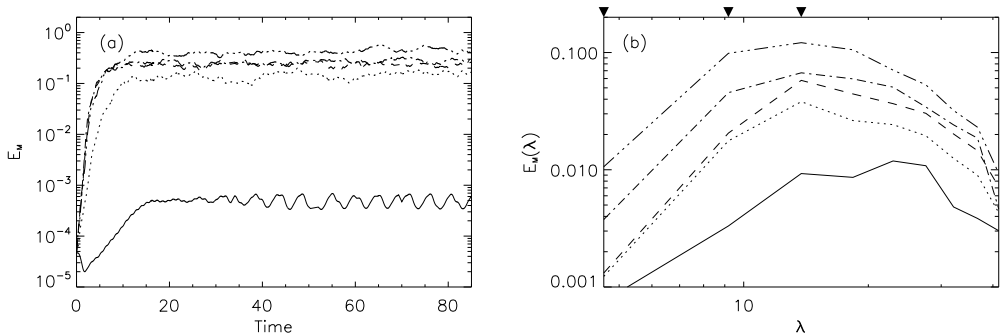}
\end{center}
\caption{(a) Magnetic energy in dynamo runs D1 (solid), D2 (dotted), D3 
(dashed), D4 (dash-dotted) and D5 (dash-triple dotted). (b) Magnetic energy 
spectrum for the same runs at late times, after non-linear saturation of 
the dynamo takes place. The arrows on top indicate the scale where mechanical 
energy is injected in each run; from left to right: D1, D2, and D3-D5. The 
magnetic energy spectrum corresponding to Run D1 has been multiplied by a 
factor of 100.}
\label{fig:trend1}
\end{figure}

The global evolution of the system is similar to what we will show in the 
remaining runs. Once the magnetic field is introduced at $t=0$, and if 
$R_m$ is large enough, the magnetic field is amplified exponentially (this 
stage is often called the ``kinematic dynamo'' regime) until the Lorentz 
force modifies the flow and non-linear saturation is reached. At late 
times, an MHD state is reached in which magnetic energy is sustained 
against Ohmic dissipation by dynamo action. In run D1, the flow is 
reminiscent of the hydrodynamic flow previously described in Figure 
\ref{fig:3Dekman}. Although the forcing is axisymmetric and purely 
toroidal, rotation generates a poloidal circulation and as a result the 
flow points outwards in the equatorial plane, and inwards along the axis 
of rotation. Each hemisphere has mechanical helicity of opposite signs, 
while the net mechanical helicity of the system fluctuates around zero. 
The magnetic field seems to be sustained by an $\alpha$-$\Omega$ mechanism, 
where the differential rotation is sustained by the mechanical forcing 
and the $\alpha$-effect is given by the Ekman-like circulation. Magnetic 
energy is concentrated in the center of the sphere, where the flow has 
a stagnation point.

\begin{figure}
\begin{center}\includegraphics[width=5cm]{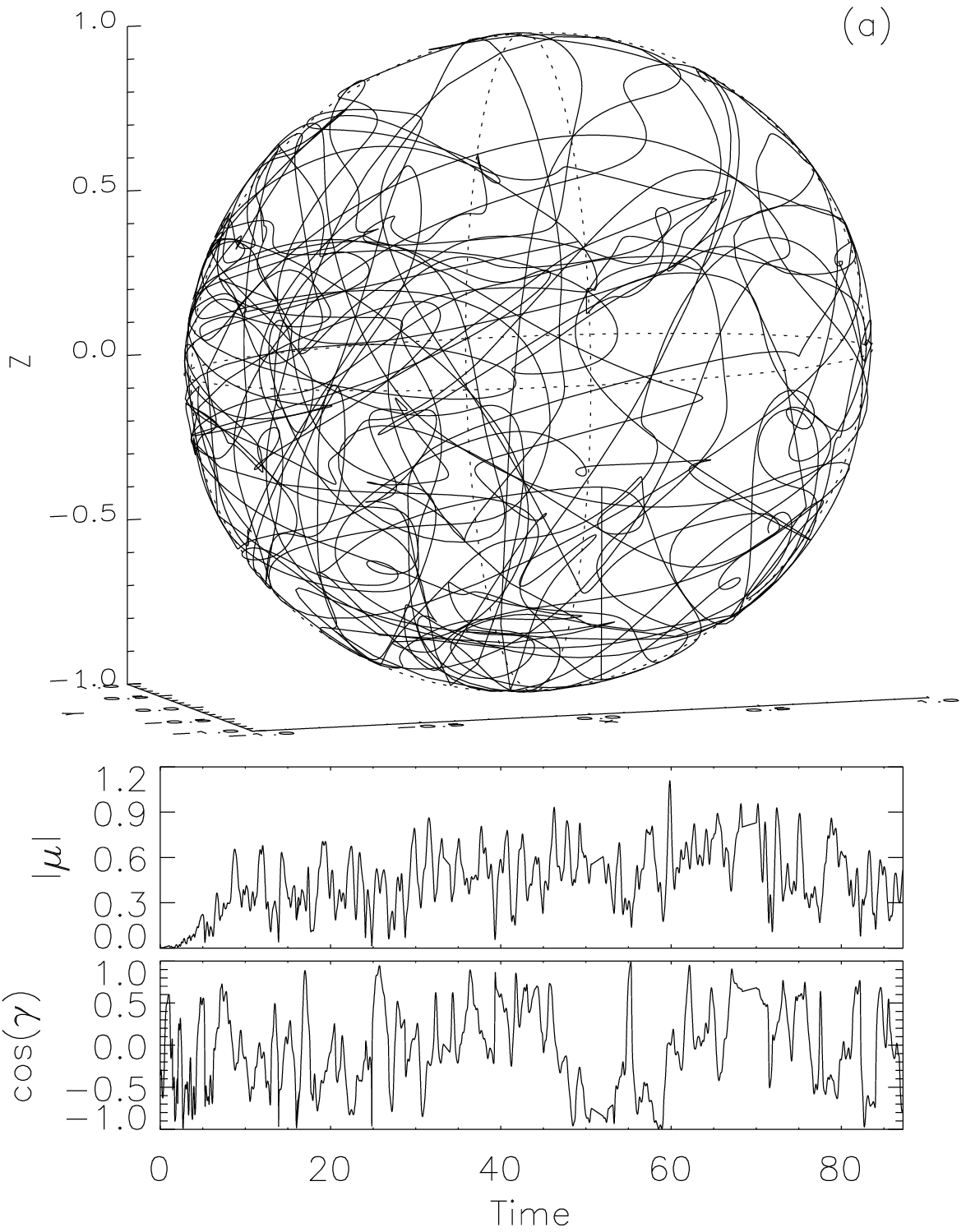}
              \includegraphics[width=5cm]{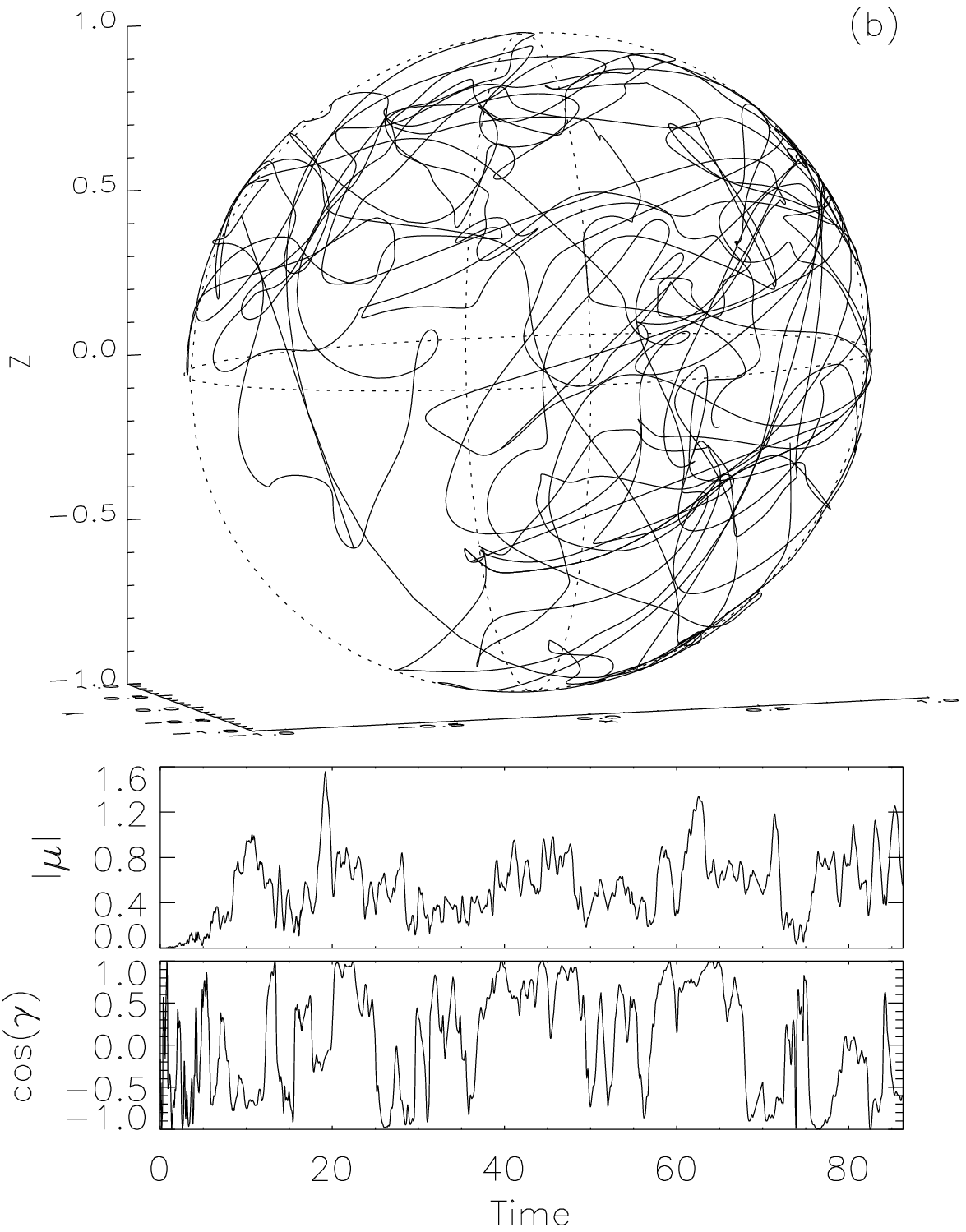}
              \includegraphics[width=5cm]{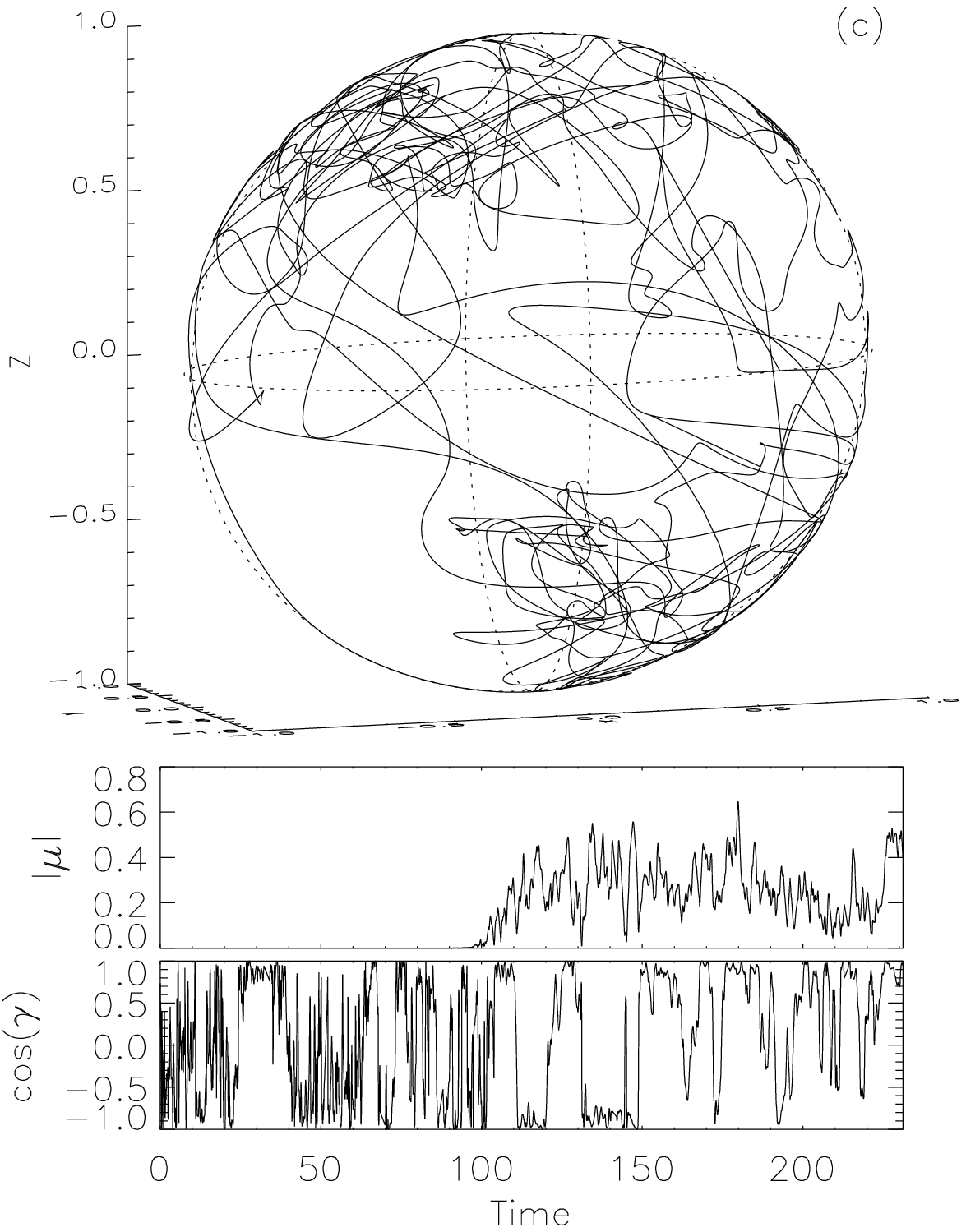}
\end{center}
\caption{Left: (a) Trace of the dipole moment on the surface of the unit 
sphere (above), amplitude of the dipole moment (middle), and cosine of the 
angle between the dipole moment and the axis of rotation as a function of 
time, for run D3. (b) Same quantities for run D5. (c) Same quantities for 
run D7.}
\label{fig:trend2}
\end{figure}

A considerably stronger dynamo than D1 is represented in Run D2, where 
in the excitation function (\ref{eq:force1}) we choose $q=l=2$ and 
$|\xi^f_{q,l,0}|=1.1$. Again, $A=B=1$, $\nu=\eta=0.002$, and $\Omega=2$. 
The magnetic moment rises from zero, and attains a typical magnitude 
of $|\vmu|\approx 0.5$, about 100 times larger than in D1. The Reynolds 
numbers are $R_e = R_m \approx 500$, and the ratio of magnetic to kinetic 
energy oscillates around $0.15$. The cosine of $\gamma$ (the angle 
between $\vmu$ and $\vOmega$) oscillates wildly in time, and the 
orientation of the dipole shows no preferred direction. The only 
difference between run D1 and D2 is the change in the forcing scale, and 
the result seems to indicate a separation of scales between the forcing 
and the largest scale in the system helps the dynamo, as indicated by 
the larger ratio $E_M/E_V$ in run D2, and as also reported before in 
simulations with periodic boundary conditions 
\cite{Brandenburg01,Gomez04,Mininni05a}. In all these runs, the largest 
available scale is fixed and given by the inverse of the smallest $|\lambda|$ 
(corresponding to $\lambda_{\pm1,1}$) and determined by the radius of the 
sphere ($R=1$), while the separation between this scale and the forcing 
scale is controlled by the values of $q$ and $l$ in the forcing function 
(see Table \ref{table:dynamo}). The largest the values of $|q|$ and $l$, 
the smallest the scale where mechanical energy is injected.

Several runs were done (see e.g. the runs D1 to D5 in Table 
\ref{table:dynamo}) in which the forcing was gradually moved to smaller 
scales (from $q=l=1$ in D1 to $q=l=3$ in D3), and in which the rotation 
rate was progressively increased (from $\Omega = 2$ in runs D1-D3 to 
$\Omega = 8$ in D5). As these changes were made, the amplitude of the 
forced modes $|\xi^f_{q,l,m}|$ in equation (\ref{eq:force1}) had to be 
increased in order to reach a statistically steady state with r.m.s. 
velocities of order one before the magnetic field was introduced (their 
amplitudes were $0.4$, $1.1$, $1.6$, $2.2$, and $3.6$, from run D1 to D5). 
The reason for this can be understood as follows: the Coriolis force in 
equation (\ref{eq:momentum}) acts as a restoring force that opposes the 
growth of perturbations. This is also the reason why this system can 
sustain waves, as was shown in Section \ref{sec:hdresults}. In all these 
runs, $A$, $B$, $\nu$, $\eta$, and the angle of inclination $\alpha$ were 
kept the same. As will be seen from Table 1, all the forcing was non-helical.

\begin{figure}
\begin{center}\includegraphics[width=8cm]{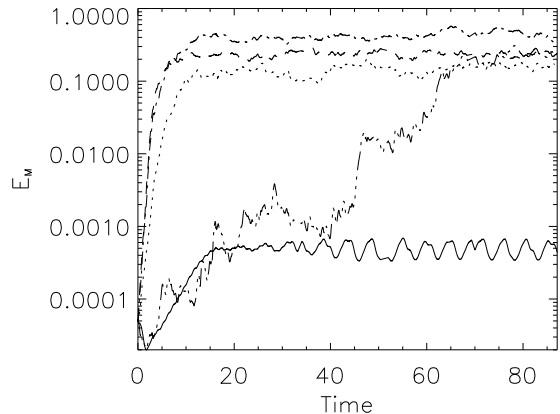}
\end{center}
\caption{Magnetic energy in dynamo runs D1 (solid), D2 (dotted), D3 
(dashed), D5 (dash-dotted) and D7 (dash-triple dotted). Note the 
intermittent growth of magnetic energy at early times in Run D7.}
\label{fig:intermittency}
\end{figure}

All five runs are considered to have been able to resolve the Ekman layers 
that developed, but they would likely not have been resolved at higher 
values of $\Omega$. Figure \ref{fig:trend1} shows the general trend 
resulting from the smaller scale forcing and increased rotation. Figure 
\ref{fig:trend1}(a) shows a logarithmic-linear plot of the total magnetic 
energy versus time for runs D1 to D5. The  five runs showed increasingly 
large growth rate, a higher saturation level of $E_M/E_V$, and increasing 
$|\vmu|$. In Run D5, the ultimate ratio of $E_M$ to $E_V$ was about $0.4$ 
and $|\vmu|$ was close to unity. Figure \ref{fig:trend1}(b) shows magnetic 
energy spectra for runs D1 to D5, with decreasing Rossby number $R_O$ 
(based on the radius of the sphere) of $0.5$ (runs D1 to D3), $0.25$ 
(D4), and $0.125$ (D5). It will be seen that there develops a 
small excess of magnetic energy in scales larger than the forcing scale 
with decreasing Rossby numbers. Note the development of a ``bump'' 
in the magnetic energy spectrum at $\lambda \approx 9$ in runs D4 and 
D5.

\begin{figure}
\begin{center}\includegraphics[width=15cm]{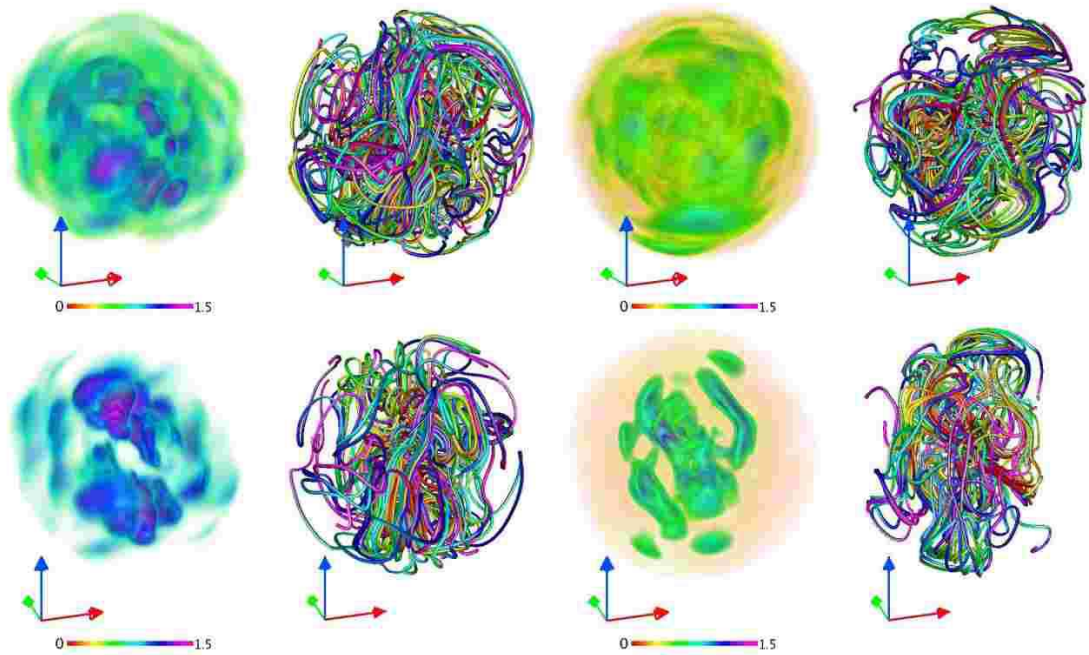}
\end{center}
\caption{Above: kinetic energy density and velocity field lines at late 
times in Run D5 (left), and magnetic energy density and magnetic field 
files at the same time for the same run. Below: same quantities for Run 
D7. Note the development of anisotropies in the presence of large 
$\vOmega$ in this run.}
\label{fig:D5D7}
\end{figure}

A second trend is indicated in Figure \ref{fig:trend2}: namely the 
orientation of the dipole moment onto the axis of rotation seems less 
erratic with decreasing Rossby and Ekman numbers. That is, there are more 
eddy turnover times (in units of $R/U$) between the near reversals 
as $R_O$ and $E_K$ are decreased, and the projection of the dipole 
moment on the unit sphere gets more localized around the two ``poles'' 
defined by the axis of rotation. We borrow here the term ``reversal'' 
from the palaeomagnetic record, where during a reversal the orientation 
of the dipole moment changes about $180^\circ$ and its amplitude 
decreases, in opposition to an ``excursion'' in which the direction and 
the orientation of the dipole moment changes in a short period of time 
without resulting in a full reversal \cite{Valet05}.

To verify this behavior, in Runs D6, D7, and D8 we successively 
decreased $R_O$ and $E_K$ while keeping the other parameters constant. 
At the present resolution, we could not decrease the values of $R_O$ 
and $E_K$ below the values for run D8 while keeping the boundary layer 
well resolved. In geodynamo simulations, a similar effect was 
reported, and it was noted that the behavior of the dipole moment was 
controlled by the amplitude of the Rossby number, independently of the 
values of the Ekman and Rayleigh numbers \cite{Christensen06}.

Figure \ref{fig:trend2} shows the trace of the dipole moment $\vmu$ 
on the surface of the unit sphere, its amplitude, and the cosine of the 
angle between $\vmu$ and $\vOmega$ for runs D3, D5, and D7. While in Run 
D3 the trace of $\vmu$ fills the entire surface of the sphere, as 
$R_O$ and $E_K$ are decreased $\vmu$ seems to fluctuate around two 
regions in opposite sides of the sphere. These regions get more localized 
with decreasing $R_O$ and $E_K$. Also, the time between excursions of 
$\vmu$ outside these regions gets larger, as shown by $\cos(\gamma)$. 
In Run D7, after a transient that finishes at $t \approx 80$, 
$\cos(\gamma)$ stays at $1$ or $-1$ for $\approx 20$ turnover times 
before changing sign rapidly. It is also worth noticing that as 
$R_O$ and $E_K$ decrease, the time it takes the system to develop a 
dipole moment of order one gets larger (see the evolution of $|\vmu|$ at 
early times in Figure \ref{fig:trend2}). This is also observed in the 
time evolution of the magnetic energy (see Figure \ref{fig:intermittency}). 
Instead of having an exponential growth of $E_M$ at early times as in Runs 
D1 to D5, Runs D6 to D8 show a more intermittent behavior: magnetic energy 
grows in rapid ``bursts'' and stays around that value until a new burst 
increases the magnetic energy again. In Run D7, this process saturates 
around $t\approx 80$ and no further change in the average value of $E_M$ 
is observed. It is possible that the slow-down during the kinematic 
dynamo regime is a consequence of a quasi-two-dimensionalization of the 
flow by the high rotation rate; this remains to be investigated fully.

\begin{figure}
\begin{center}\includegraphics[width=15cm]{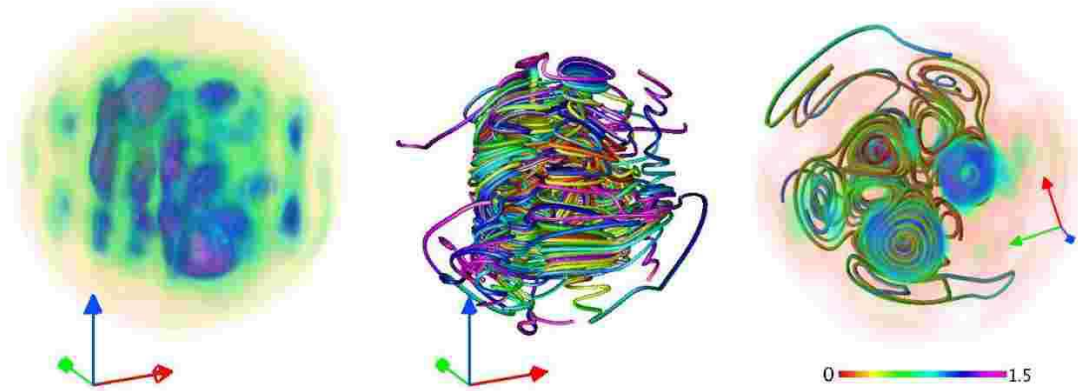}
\end{center}
\caption{Kinetic energy density (left), velocity field lines (middle), and 
view from top of the kinetic energy density superposed with velocity 
field lines (right) in run D10, before magnetic energy is introduced. Note 
the columnar structures in the velocity field aligned along $\vOmega$ 
(in the $z$ direction).}
\label{fig:columns}
\end{figure}

Figure \ref{fig:D5D7} shows visualizations of energy densities and 
field lines at late times in Runs D5 and D7. The development of 
anisotropies in the velocity and magnetic fields can be observed 
in run D7, which has the highest rotation rate attained of $\Omega=16$. 
Indeed, as $\Omega$ is increased the velocity field shows a tendency 
to develop columns, with mechanical energy concentrated in cylindrical 
structures aligned along $\vOmega$ and with a larger component of $v_\phi$ 
than of $v_z$. The velocity field in these column is helical, although in 
general the total mechanical helicity of the flow fluctuates in time 
around zero. These structures are observed before the magnetic field is 
introduced (although they persist as the magnetic energy grows) and seem 
to be the result of the Taylor-Proudman effect (see e.g. 
\cite{Taylor23,Davies72}). It is a trend observed through runs D5 to D10 
(see Figure \ref{fig:columns} for an example).

Run D9 experiments with lowering the angle between the forcing's axis 
of symmetry and the axis of rotation to $\alpha = 20^\circ$, and the 
dipole becomes more difficult to excite (as evidenced by a smaller growth 
rate in the kinematic dynamo regime). Indeed, dynamo action could not be 
excited below $15^\circ$ for the values of Reynolds numbers explored, 
as the resulting driven flow approaches axisymmetry.

\begin{figure}
\begin{center}\includegraphics[width=7.2cm]{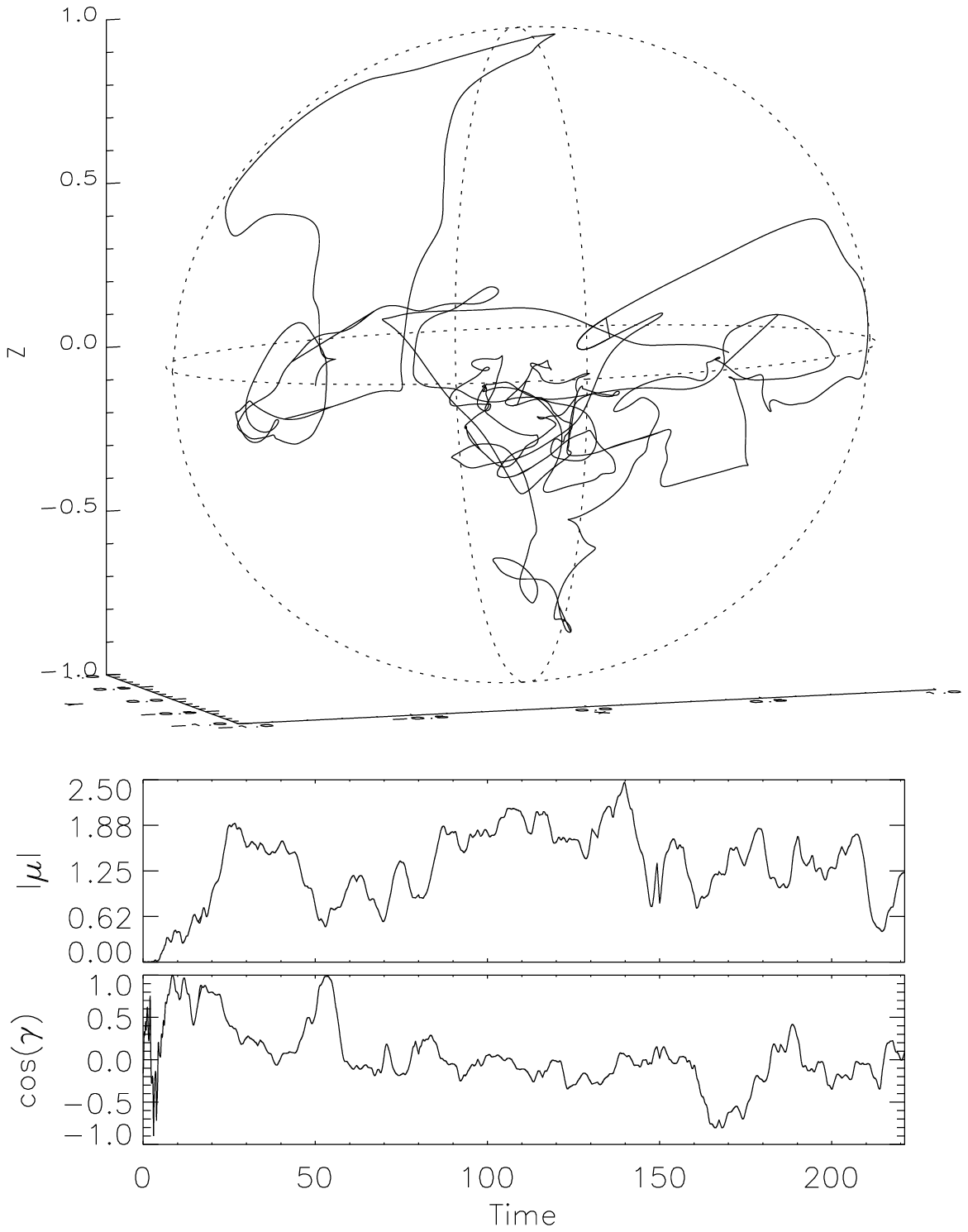}
              \includegraphics[width=7.2cm]{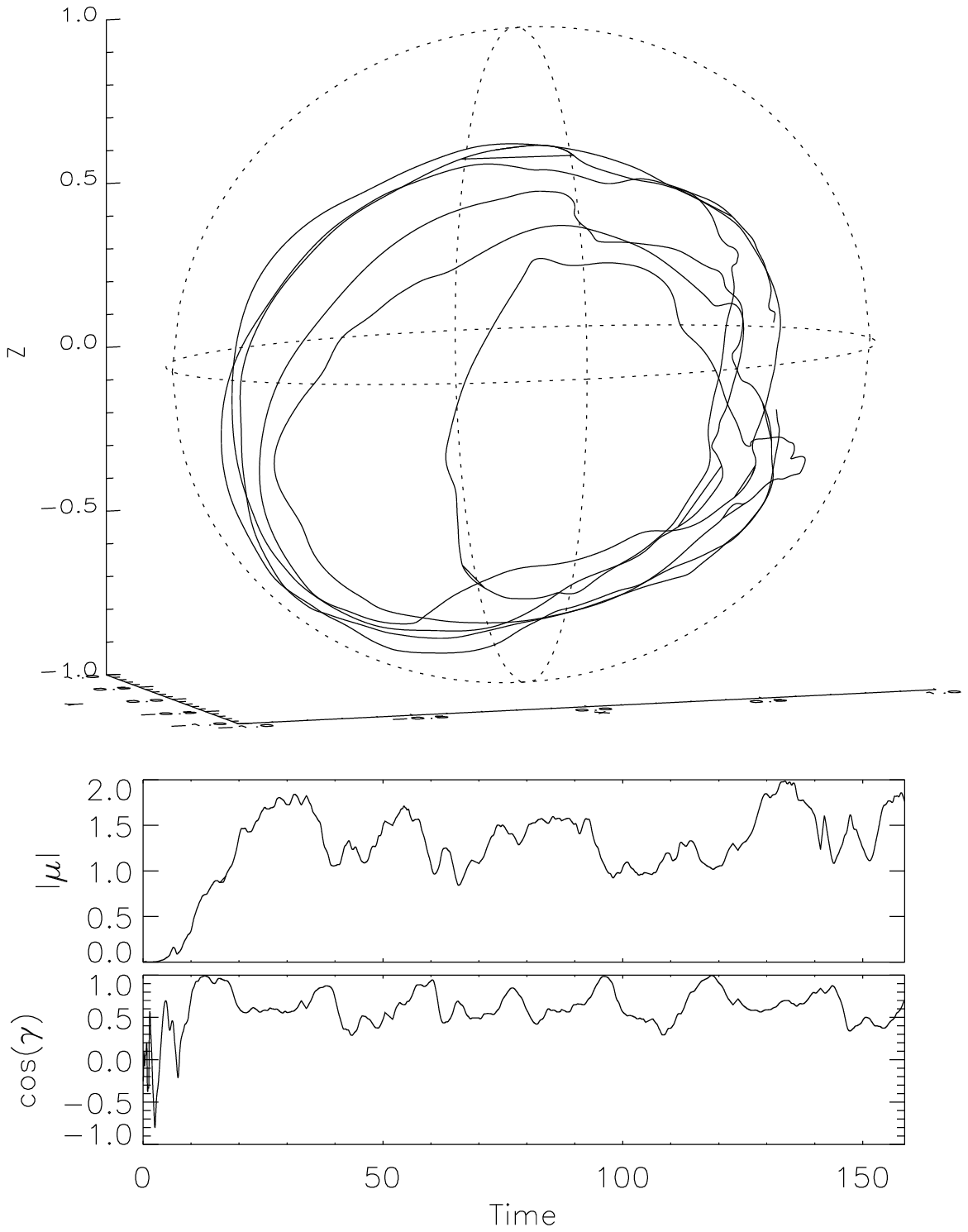}
\end{center}
\caption{(a) Trace of the dipole moment on the surface of the unit sphere 
(above) and amplitude of the dipole moment as a function of time (below) 
for Run D11. (b) Same quantities for run D12.}
\label{fig:diphel}
\end{figure}

Run D10 is actually part of a set of experiments (Runs D10 to D12; again, 
see Table \ref{table:dynamo}) with forcing functions that are 
non-axisymmetric, and which may also inject mechanical helicity (Runs 
D11 and D12). Run D10, having no net mechanical helicity, shows no 
big differences in the evolution of global quantities from the previously 
discussed runs. The dipole develops but its orientation wanders randomly, 
with some preferred orientation perpendicular to $\vOmega$. In this case, 
axisymmetry is broken by the forcing directly instead of by a non-zero 
angle between an axisymmetry forcing and the axis of rotation.

Runs D11 and D12 have non-axisymmetric forcing that injects mechanical 
helicity. The forcing for these runs is given by equation 
(\ref{eq:force2}) with coefficients
\begin{equation}
\xi^f_{3,3,0} = 5 \xi^f_{-3,3,0} = F_0 , \;\; 
\xi^f_{3,3,0<m\le 3} = 5 \xi^f_{-3,3,0<m\le 3} = F_0 (1+i) ,
\end{equation}
with $F_0 = 1.7$. In the presence of net helicity, dynamo excitation 
suddenly becomes much easier (as evidenced by a much larger growth rate 
of magnetic energy during the kinematic regime), and the ultimate 
saturation occurs at $E_M/E_V \approx 2$: more magnetic than kinetic, 
with magnetic helicity, having sign opposite that of the mechanical 
helicity inversely cascading to the large scales. As a result, the system 
is dominated by a helical magnetic field at the largest available scale. 

An interesting qualitative argument from mean field theory 
\cite{Steenbeck66,Krause} (which assumes large scale separations and 
often some form of periodic boundary conditions, as do all 
``$\alpha$-effect'' calculations) can be seen to anticipate this result 
as follows. From mean field theory , the induction equation for the mean 
magnetic field $\overline{\bf B}$ (assuming there is no mean flow 
$\overline{\bf U}$) is 
\begin{equation}
\frac{\partial \overline{\bf B}}{\partial t} = 
    \alpha \nabla \times \overline{\bf B} + \beta \nabla^2 \overline{\bf B} .
\label{eq:mean}
\end{equation}
Here, $\alpha$ is proportional to minus the kinetic helicity of the 
flow \cite{Steenbeck66,Pouquet76,Krause}, and $\beta$ is a turbulent 
magnetic diffusivity. Dotting equation (\ref{eq:mean}) with the mean 
vector potential $\overline{\bf A}$ (such as 
$\overline{\bf B} = \nabla \times \overline{\bf A}$) and integrating 
over volume, an equation for the evolution of the mean magnetic helicity 
$\overline{H_M}$ is obtained,
\begin{equation}
\frac{d \overline{H_M}}{d t} = 
    \alpha \overline{E_M} + \beta \nabla^2 \overline{H_J} ,
\end{equation}
where $\overline{E_M}$ is the mean magnetic energy and $\overline{H_J}$ 
is the mean current helicity. As a result, if magnetic diffusion is 
neglected, the dynamo process injects into the mean (large) scales 
magnetic helicity of opposite sign than the kinetic helicity, and in 
the small scales magnetic helicity of the same sign. This effect has 
been observed before in numerical dynamo simulations with periodic 
boundary conditions \cite{Brandenburg01,Gomez04}. The large scale magnetic 
helicity then inverse-cascades to the largest available scale in the 
system, while the small scale magnetic helicity is transferred to smaller 
scales where it is dissipated \cite{Alexakis06}. As a result, at late 
times the system is dominated by a large scale magnetic field with 
magnetic helicity of opposite sign to that of the kinetic helicity 
injected by the forcing.

\begin{figure}
\begin{center}\includegraphics[width=15cm]{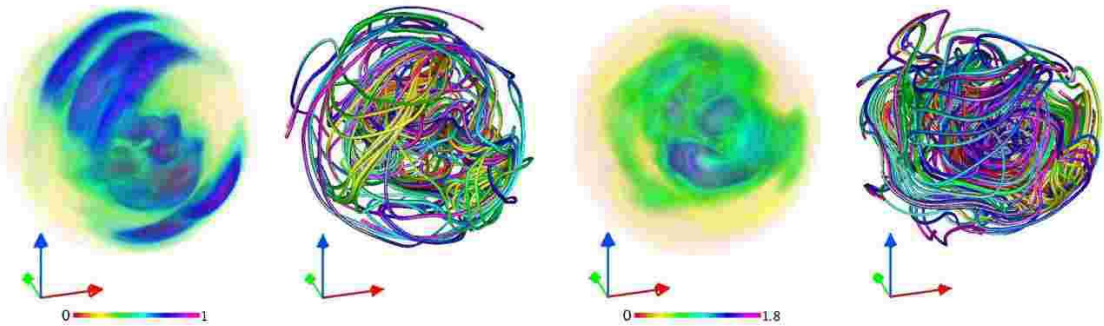}
\end{center}
\caption{Kinetic energy density and velocity field lines (left), and 
magnetic energy density and magnetic field lines (right) at the same 
time in the saturated state of run D12. The axes are aligned as in 
Figure \ref{fig:diphel}.}
\label{fig:D123D}
\end{figure}

As seen in Figure \ref{fig:diphel}, the dipolar orientation in D11 
seems to have a preference for being perpendicular to $\vOmega$. For 
run D12, whose forcing function differs in its orientation to $\vOmega$ 
by $90^\circ$, the dipole orientation seems to remain in a single 
hemisphere (see Figure \ref{fig:diphel}) and its tip precesses about 
$\vOmega$. Figure \ref{fig:D123D} shows energy density and field lines 
in the saturated steady state of run D12. The axes are aligned as in 
Figure \ref{fig:diphel}. As is often the case with helical flows, the 
geometry of the flow is more complex than in the non-helical runs. These 
runs in which a net sign of mechanical helicity is sustained by the 
mechanical forcing were continued for several thousands turnover times, 
and no reversals of the dipole moment were observed. The dipole moment 
seems to fluctuate around a preferred orientation with only short 
excursions of the dipole to the opposite hemisphere.

\section{Discussion and future directions \label{sec:future}}
By solving the mechanically-forced MHD equations inside a rotating 
conducting spherical boundary, we have found that a bewilderingly 
wide variety of dynamo behavior is possible for a magnetic Prandtl 
number of unity. The behavior is sensitive to mechanical and magnetic 
Reynolds numbers, to rotation rate, and more indirectly to the Ekman 
number. It is also sensitive to the geometry and strength of 
the forcing functions and their relation to the axis of rotation. 
We have only begun to explore this multidimensional parameter space. 
Note we have not made much effort to tailor the forcing functions we 
have chosen to model what the mechanical flows in planetary cores or 
stellar convective regions might be. Rather, we have been exploring 
dynamo behavior in the abstract, and feel somewhat overwhelmed by the 
variety of dynamo behavior that has been found. In this light, 
our attempt should be consider as an extension of dynamo simulations 
in periodic boundary conditions 
\cite{Meneguzzi81,Brandenburg01,Gomez04,Brandenburg05,Ponty05,Mininni05b} 
to consider the effect of boundaries and rotation, and not be compared 
with explorations of the space of parameters in realistic geodynamo 
simulations \cite{Glatzmaier95,Zhang00,Roberts01,Kono02,Christensen06}.

What has become clear is that the wholly spectral methods we are using, 
while accurate, do not scale well into the parameter regimes of planetary 
and astrophysical dynamos, which involve many orders of magnitude between 
the largest length scales in the flows and the Ekman or dissipation scales 
that also play a role in the process. This limitation afflicts all numerical 
attempts to explore planetary and stellar dynamos, particularly in view of 
the low magnetic Prandtl numbers that are expected to prevail there, 
in simulations \cite{Ponty05,Mininni05b}, and in liquid-sodium experiments 
\cite{Gailitis01,Steiglitz01,Petrelis03,Sisan03,Spence05}. But the rapid 
multiplication of the number of terms to compute in the convolution sums 
in equations (\ref{eq:CK1}) and (\ref{eq:CK2}) provides a rather 
intractable limitation on efforts to use our code without modification 
at higher Reynolds numbers than the few hundred we have explored here. 
What seems to be called for is an exploration of the possibilities of 
using fast transforms (for spherical harmonics 
\cite{Driscoll94,Mohlenkamp99} and possibly for spherical Bessel functions 
\cite{Sharafeddin92,Cree93})to turn the code into a pseudospectral one 
in which the nonlinear terms are computed in configuration space rather 
than spectral space, as is commonly done for rectangular periodic 
boundary conditions \cite{Canuto} which would increase available 
resolution by many orders of magnitude. Our future investigations will 
explore this possibility.

We also intend to replace the conducting boundary by an insulating but 
mechanically-impenetrable one that will permit protrusion of the magnetic 
field into the vacuum region surrounding the shell (see e.g. \cite{Kono02}). 
It may be that forcing the magnetic field lines to return to the interior 
each time they get near the shell that is now playing a dynamical role in 
what we are seeing would be different in the case of an insulating shell. 
This raises some conceptual difficulties, since the problem of matching 
MHD fields on to vacuum electromagnetic ones has not been thoroughly 
solved. For example, one approximation that has been used has been to 
match magnetic fields at 
a spherical surface to magnetostatic ones outside, which involve a magnetic 
field that is derivable from a scalar potential and for which there is no 
electric field. However, Maxwell's equations tell us that the tangential 
electric field must be continuous at any interface, and there is no reason 
why this tangential electric field should vanish or even be ``small'' 
immediately inside an insulating boundary of a conducting magnetofluid. 
The magnetostatic approximation may be the best we can do, but it 
would be desirable to have more justification for it than we presently 
have.

Finally, we need to devote attention to the geometry and strength of 
the forcing functions that are being employed and to study the effect of 
different forcing functions in dynamo action. Mechanical processes, 
convective and otherwise, are believed to power the dynamo in cases of 
geophysical or astrophysical interest. In the simulations discussed, we 
have no explanation, for example, why the ``columns'' or columnar vortices 
aligned along $\vOmega$ form in the velocity field in runs toward the 
end of Table \ref{table:dynamo}. It is clear the physical effect that 
is triggering their formation is the rotation alone, perhaps through a 
two-dimensionalization of the flow via the Taylor-Proudman effect 
\cite{Taylor23,Davies72,Acheson}. This is a different process than the 
conventional explanation for the formation of columns in planetary interiors, 
which involves both thermal convection and rotation \cite{Busse70}, since 
thermal convection is completely absent in our incompressible MHD 
formulation.

\ack
Computer time was provided by NCAR. NSF grants CMG-0327888 at NCAR, 
ATM-0327533 at Dartmouth, and AST-0507760 at Cornell supported this work 
in part and are gratefully acknowledged.

%\bibliography{ms}

\end{document}